\documentclass[iop]{emulateapj}
\pdfoutput=1
\usepackage[utf8]{inputenc}
\usepackage{apjfonts}
\usepackage{multirow}

\usepackage{graphicx}
\usepackage{epstopdf}

\newcommand{\myshorttitle}{Wide-Field Near-IR Imaging of M31}
\newcommand{\myshortauthors}{Sick et al.}

\usepackage{color}
\usepackage[dvipsnames]{xcolor}
\usepackage[pdfauthor={\myshortauthors},pdftitle={\myshorttitle},colorlinks=true,citecolor=blue,linkcolor=blue,urlcolor=blue]{hyperref}
\usepackage{url}

\usepackage{amssymb}
\usepackage{amsmath}

\usepackage{natbib}
\newcommand{\ie}{\textit{i.e.,~}}
\newcommand{\eg}{\textit{e.g.,~}}

\newcommand{\vect}[1]{\boldsymbol{#1}} \newcommand{\sw}[1]{\textit{#1}} \newcommand{\sn}{\ensuremath{S/N}} 
\newcommand{\iiwione}{\sw{`I`iwi 1.0}}
\newcommand{\androids}{\textsc{androids}}
   \newcommand{\Fig}[1]{Fig.~\ref{fig:#1}}  \newcommand{\Eq}[1]{Eq.~\ref{eq:#1}}  \newcommand{\Tab}[1]{Table~\ref{tab:#1}}  \newcommand{\Sec}[1]{\S\ref{sec:#1}}  

\shorttitle{\myshorttitle}
\shortauthors{\myshortauthors}

\begin{document}
    \slugcomment{Submitted to AJ.}
\title{Andromeda Optical and Infrared Disk Survey \sc{i}.\\ New Insights in Wide-Field Near-IR Surface Photometry}
\author{Jonathan Sick, Stéphane Courteau}
\affil{Department of Physics, Engineering Physics \& Astronomy, Queen's University, Kingston, Ontario, K7L 3N6 Canada}
\email{jsick@astro.queensu.ca}
\author{Jean-Charles Cuillandre}
\affil{Canada-France-Hawaii Telescope Corp., Kamuela, HI 96743, USA}
\author{Michael McDonald}
\affil{Kavli Institute for Astrophysics and Space Research, MIT, Cambridge, MA 02139, USA}
\author{Roelof de Jong}
\affil{Astrophysikalisches Institut Potsdam (AIP), An der Sternwarte 16, 14482 Potsdam, Germany}
\and
\author{Brent Tully}
\affil{Institute for Astronomy, University of Hawaii, 2680 Woodlawn Drive, Honolulu, HI, USA}

\begin{abstract}
We present wide-field near-infrared $J$ and $K_s$ images of the Andromeda Galaxy taken with WIRCam on the Canada-France-Hawaii Telescope (CFHT) as part of the Andromeda Optical and Infrared Disk Survey (\androids).
This data set allows simultaneous observations of resolved stars and NIR surface brightness across M31's \emph{entire} bulge and disk (within $R=22$~kpc), permitting a direct test of the stellar composition of near-infrared light in a nearby galaxy.
Our survey complements the similar Panchromatic Hubble Andromeda Treasury survey by covering M31's entire disk, rather than a single quadrant, at similar wavelengths, albeit with lower spatial resolution.
The primary concern of this work is the development of NIR observation and reduction methods to recover a uniform surface brightness map across the $3\arcdeg \times 1\arcdeg$ disk of M31.
This necessitates sky-target nodding across 27 WIRCam fields.
Two sky-target nodding strategies were tested, and we find that strictly minimizing sky sampling latency does not maximize sky subtraction accuracy, which is at best 2\% of the sky level.
The mean surface brightness difference between blocks in our mosaic can be reduced from 1\% to 0.1\% of the sky brightness by introducing scalar sky offsets to each image.
We test the popular \sw{Montage} package, and also develop an independent method of estimating sky offsets using simplex optimization; we show these two optimization schemes to differ by up to 0.5~mag~arcsec$^{-2}$ in the outer disk.
We find that planar sky offsets are not acceptable for subtracting residual backgrounds across WIRCam fields that are much smaller than the mosaic area.
The true surface brightness of M31 can be known to within a statistical zeropoint of 0.15\% of the sky level (0.2 mag arcsec$^{-2}$ uncertainty at $R=15$~kpc). 
We also find that the surface brightness across individual WIRCam frames is limited by both WIRCam flat field evolution and residual sky background shapes.
To overcome flat field variability of order 1\% over 30 minutes, we find that WIRCam data should be calibrated with real-time sky flats.
Due either to atmospheric or instrumental variations, the individual WIRCam frames have typical residual shapes with amplitudes of 0.2\% of the sky after real-time flat fielding and median sky subtraction.
We present our WIRCam reduction pipeline and performance analysis here as a template for future near-infrared observers needing wide-area surface brightness maps with sky-target nodding, and give specific recommendations for improving photometry of all CFHT/WIRCam programs.
\end{abstract}
\section{Introduction}
\label{sec:intro}

Thanks largely to its proximity, kinship with the Milky Way, and being an ideal foil for modern galaxy formation models, the Andromeda galaxy has been the focus of numerous investigations of galaxy structure \citep{Ibata:2005,Irwin:2005,McConnachie:2009,Courteau:2011}
and stellar populations \citep{Williams:2002,Worthey:2005,Saglia:2010}.
The shapes, ages, kinematics and relative fraction of galaxy components (bulge, disk, halo) in large spiral galaxies like our own reveal precious information about their formation, accretion, and merging histories \citep[see the review of][]{Kormendy:2004}.

Unfortunately, the fundamental task of disentangling galaxy components---which typically involves light profile decompositions, colour gradients and mass modelling---remains non-trivial.
Stellar population studies of spiral galaxies are thwarted by a three-fold degeneracy between stellar age ($A$), metallicity ($Z$) and ISM dust that can only be lifted by combining, at the very least, optical and infrared images with realistic dust models \citep{de-Jong:1996b,MacArthur:2004,Pforr:2012}.
Likewise, mass models of spiral galaxies suffer a degeneracy between the stellar $M/L$ and dark halo parameters that requires, in addition to an extended rotation curve, deep and accurate multi-band imaging to uniquely constrain the stellar $M/L$ ratio \citep{Dutton:2005}. 

Compounding these challenges are fundamental uncertainties in modern stellar population synthesis, particularly uncertainties in the interpretation of near-infrared (NIR) light.
In the Galaxy and Mass Assembly (GAMA) survey, \cite{Taylor:2011} combined SDSS and UKIRT $ugrizYJHK$ photometry of galaxies to show that optical-NIR SEDs yield unreliable population synthesis fits compared to optical-only SED fits.
This failure is largely attributable to inadequate stellar population synthesis recipes for NIR bands and naive parameterization of star formation histories.

First, spectral energy distribution (SED) fitting often relies on simplistic star formation history (SFH) parameterizations.
Because NIR colours lift age-metallicity-dust degeneracies, modelling of NIR bands may require additional sophistication, namely composite star formation and metal enrichment histories.
The appropriate form of SFH models cannot be constrained from the integrated light of galaxies alone (as is typically attempted); resolved colour-magnitude diagrams (CMDs) are both more effective and, in fact, essential at deriving non-parametric stellar population histories. 

Second, the NIR light is dominated by thermally-pulsating AGB (TP-AGB) stars from intermediate-aged stellar populations.
Modelling TP-AGB stars is most challenging due to their complex dredge-up cycles that change surface chemistry and temperature (the M- to C-type transition), and circumstellar winds that further perturb an AGB star's location in the CMD.
A proper calibration of NIR stellar population synthesis models \cite[\eg][Charlot \& Bruzual in prep.]{Maraston:2005} will yield a 30\%--50\% improvement in the estimation of stellar masses and ages of high redshift systems \citep[\eg][]{Maraston:2006,Bruzual:2007,Conroy:2010b,Conroy:2013}.

Our remedy for both understanding the structure of M31, and more fundamentally to understand NIR stellar populations, is to survey the entire bulge and disk ($R \leq 22$~kpc) of M31 in both resolved and integrated stellar light at $J$ and $K_s$ wavelengths.
In doing so, we can directly relate a NIR stellar population's decomposition in the colour-magnitude plane to the panchromatic SED of M31.
Though such a calibration could be made with other galaxies, M31 is unique in its proximity so that even ground-based instrumentation can resolve its bright stellar population
For reference, $1\arcsec = 3.7~\mathrm{pc}$ across the disk of M31 \citep[we adopt $D_\mathrm{M31} = 785$~kpc,][]{McConnachie:2005}.

A wealth of photometric data exist for M31, however, none provide simultaneous observations of M31's resolved and integrated NIR light.
The best resource for resolved stellar populations across the entire disk of M31, to date, is the Local Group Galaxy survey \cite[LGGS,][]{Williams:2003,Massey:2006}.
This survey, although covering the $UBVRI$ wavelengths, does not extend to the $JHK$ NIR wavelengths most contentious for stellar population models.
Advancing our view of M31 to NIR wavelengths, \cite{Beaton:2007} assembled a 2.8\arcdeg\ $JHK_s$ mosaic of M31 with the 2MASS 6X program.
Those observations, the first nearly dust-free of M31's stellar content, were used by \cite{Athanassoula:2006} as evidence for a bar embedded in a classical bulge.
Beyond the bulge, the 2MASS 6X images have limited utility; sky uncertainties restrict their use to infer structural and photometric properties of the disk.
Further, the pixel scale of 1\arcsec\ and integration depth of 46.8 seconds prevent point source measurement of M31 stars in the 2MASS 6X images.
As a result, the state-of-the-art NIR view of M31 is the slightly longer 3.6 $\mu$m Spitzer/IRAC map of \cite{Barmby:2006}.
Spitzer reduces the uncertainties of sky estimation, though the pixel scale of 0\farcs86 also prevents point source measurements of individual stars in the M31 disk.

Decompositions of M31's stellar populations have been made with high-resolution Hubble Space Telescope (HST) observations \citep{Brown:2003,Brown:2006,Brown:2008}.
These authors detected an intermediate age population in the inner halo of M31, and even disentangled debris associated with the Giant Stream around M31 in six fields sampling the outer disk, halo, and Giant Stream around M31 \citep{Brown:2009a}.
Similarly, though in the near-infrared, \cite{Olsen:2006} derived star formation histories within $22\farcs5\times22\farcs5$ fields in M31's inner disk and bulge with ground-based adaptive optics observations with the Gemini ALTAIR/NIRI instrument.
Yet these pencil beam surveys cannot be construed as representative of the entire Andromeda Galaxy.
Thus the boldest step forward in understanding M31's stellar populations is coming from the Panchromatic Hubble Andromeda Treasury Survey \citep[PHAT,][]{Dalcanton:2012}.
PHAT provides wide-field coverage from M31's centre to the $10$~kpc star forming ring, a panchromatic view of stellar populations from 3000\AA--17000\AA, and resolution of stars in very crowded environments such as the bulge.
Despite its many enviable traits, PHAT only covers a single quadrant of M31, making any global conclusions about M31's stellar populations incomplete.
Further, the wavelength coverage of HST/WF3 falls short of the $2.2$~$\mu$m $K$-band, making empirical calibration of the common NIR bands used by wide-field NIR surveys ($JHK$) impossible.
Thus there is good cause, even in the era of PHAT, to revisit M31 with a ground-based survey that covers the entire M31 disk at NIR wavelengths, while using the best natural seeing in the northern hemisphere (on Mauna Kea) to resolve stars even more effectively than the previous ground-based survey of M31's disk (LGGS).

We present such a survey, the first instalment of the Andromeda Optical and Infrared Disk Survey (\androids), in this paper.
ANDROIDS uses the WIRCam instrument \citep{Puget:2004} on the Canada-France-Hawaii Telescope (CFHT), which is among the first generation of wide-field ground-based NIR detector arrays, covering a $21\farcm 5 \times 21\farcm 5$ field of view.
Indeed, the advent of detectors such as WIRCam makes such a wide-field, high-resolution survey of an object as vast as M31 possible.
The excellent natural seeing on Mauna Kea of 0\farcs 65 is sufficient for resolving giant-branch stars throughout the disk of M31.

Recovering the true NIR surface brightness map of M31 is, however, more technically challenging.
The NIR sky is $\sim 10^3\times$ brighter than the NIR surface brightness of M31 at $R=20$~kpc, demanding exceptionally careful sky background characterization.
Whereas most NIR galaxy surveys can measure the instantaneous background blank sky pixels surrounding the galaxy on a detector array, M31's size requires physically nodding of the telescope away from the galaxy by 1\arcdeg--3\arcdeg\ to sample blank sky (called Sky-Target, or ST, nodding).
That we can never observe the instantaneous sky emission on the disk of M31, but rather sample the sky at both a different location and time, introduces additional complications.
\cite{Adams:1996} clearly showed, with $9\arcdeg \times 9\arcdeg$ movies of the sky, that NIR sky emission has coherent spatial structure that moves across the sky, akin to a cirrus cloud system.
This assures that sky sampled from a sky field \emph{will not} correspond directly to the sky affecting disk observations.

Another concern is the accuracy of the surface brightness shape across individual WIRCam fields of view.
Spatial structures in the NIR sky can leave residual sky shapes in background subtracted disk images that ultimately affect our ability to produce a seamless NIR mosaic of M31.
\cite{Vaduvescu:2004} also found that detector systems themselves, in their case the (now decommissioned) CFHT-IR camera, can add a time-varying background signal whose strength may be comparable to the NIR surface brightness of the outer M31 disk.

Because such a large mosaic has never before been assembled in a ST nodding WIRCam program, we focus this contribution on engineering the best practices for this type of observing.
This includes: finding the optimal ST nodding cadence, defining the appropriate data reduction procedures for a WIRCam surface brightness reduction, and finally presenting an analysis of the surface brightness accuracy in wide-field WIRCam mosaics.

Section~\ref{sec:Observations} describes the novel observational strategies used to reduce sky subtraction uncertainties.
Section~\ref{sec:reduction} presents the image reduction pipeline; with night sky flat fielding in Section~\ref{sec:flats}, median sky subtraction in Section~\ref{sec:mediansky}, and zeropoint calibration practises in Section~\ref{sec:photocal}.
In Section~\ref{sec:scalar} we present our method for recovering the galaxy surface brightness by minimizing image-to-image differences across the mosaic, while in Section~\ref{sec:scalaranalysis} we analyze the results of this algorithm.
We estimate the systematic uncertainties in our mosaic solution in Section~\ref{sec:systematics}, where we also compare our technique to the Montage package \citep{Berriman:2008} and the Spitzer/IRAC mosaics.
In Section~\ref{sec:skyflatstability} we consider the accuracy of the surface brightness \emph{shapes} recovered by our pipeline.
Ultimately we seek the observation and reduction method that maximizes surface brightness accuracy.
Finally in Section~\ref{sec:conclusions} we summarize the uncertainty of NIR sky subtraction on the scale of M31 and outline our ideal observation and reduction method.

\section{Observations}
\label{sec:Observations}

\begin{figure}[t]
\centering
\includegraphics[width=\columnwidth]{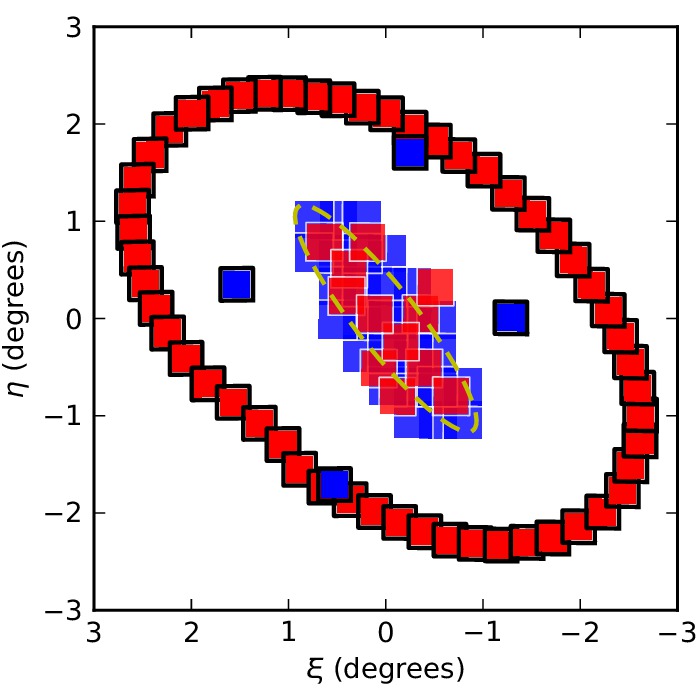}
\caption{\androids\ WIRCam field positions on M31.
Central blue fields are the 27 disk fields observed in 2007B, surrounded by 4 sky fields (blue with solid black outlines).
Red fields at center are the 12 disk fields observed in 2009B.
The red ring of 53 fields is the 2009B sky sampling ring.
The dashed yellow ellipse marks the M31 disk at $R=20$ kpc along the major axis.
Coordinates are centered on the nucleus of M31 with North up, and East is left.}
\label{fig:fieldmap}
\end{figure}

\begin{table*}[t]
\caption[Summary of WIRCam observing programs]{Summary of WIRCam observing programs.
$N_\mathrm{disk}$ is the number of WIRCam fields covering in the M31 disk in each semester (see \Fig{fieldmap}).
ST Nods with superscripts denote the number of times an observation is repeated at a given field.
$T_\mathrm{int}$ is the total integration time per disk field while $T_\mathrm{exp}$ is the integration time per WIRCam exposure.
\emph{Eff.} is the observing efficiency, or percentage of time in a program allocated to integrating the disk of M31, compared to nodding, read out and sky overheads.
$\mu_\mathrm{sky}$ gives the range (min-max) of sky surface brightnesses seen in each band, for each semester.
PSF reports the distribution seeing as measured from the full width at half maximum (FWHM) of stellar point spread functions in the uncrowded sky images.}
\label{tab:obssummary}
    
    \centering
    \begin{tabular}{lllllllllll}
        & & & & $\frac{T_\mathrm{int}}{\mathrm{field}}$ & $T_\mathrm{exp}$ & Eff. & $\mu_\mathrm{sky}$ & \multicolumn{3}{c}{PSF FWHM (arcsec)} \\ \cline{9-11}
    Semester & Band & $N_\mathrm{disk}$ & ST Nods & (min) &  (s) &  (\%) & (mag/arcsec$^2$) & 25th  & 50th & 75th \\
    \hline
    \multirow{2}{*}{2007B} & $J$ & \multirow{2}{*}{27} & [S$^3$T$^8$]$^{2}$S$^3$ & 12.5 & 47 & 49 & (15.4, 16.7) & 0.68 & 0.75 & 0.84 \\
     & $K_s$ &  & [S$^5$T$^{13}$]${^2}$S$^5$ & 10.8 & 25 & 42 & (13.4, 14.2) & 0.60 &  0.65 & 0.73 \\
     \hline
     \multirow{2}{*}{2009B} & $J$ & \multirow{2}{*}{12} & \multirow{2}{*}{[ST$^2$]$^{20}$S} & \multirow{2}{*}{13.3} & \multirow{2}{*}{20} & \multirow{2}{*}{26} & (15.0, 16.5) & 0.61 & 0.69 & 0.83 \\
                            & $K_s$ & & & & & & (13.4, 14.3) & 0.60 & 0.66 & 0.76 \\
    \end{tabular}
\end{table*}

The Andromeda Galaxy (M31) was observed in the NIR using the WIRCam instrument, mounted to the 3.6-meter Canada-France-Hawaii Telescope (CFHT), at the summit of Mauna Kea in Hawaii.
Observations were carried out in the NIR $J$ ($\lambda_0 \sim 1.2 \mu\mathrm{m}$) and $K_s$ ($\lambda_0 \sim 2.2 \mu\mathrm{m}$) bands.

WIRCam itself is an array of four HgCdTe HAWAII-RG2 detectors \citep{Puget:2004}.
Each detector comprises $2048\times 2048$ pixels, with a scale of 0\farcs 3~pix$^{-1}$.
This pixel scale critically samples the typical seeing of 0\farcs 65~at CFHT\@.
WIRCam's detectors are arranged in a $2\times 2$ grid with 45\arcsec~gaps, so that the entire instrument covers $21.5\arcmin \times 21.5\arcmin$ of sky.
It is truly the advent of NIR focal plane arrays, like WIRCam, that have enabled wide-field studies of M31 in the NIR.

The \androids\ WIRCam survey is designed to simultaneously resolve stars and recover the integrated surface brightness of the entire M31 disk.
As discussed in \Sec{intro}, NIR observations require frequent monitoring of the sky background.
\cite{Vaduvescu:2004} found, for example, that the NIR sky intensity can vary by 0.5\% per minute; yet the low surface brightness of M31's NIR disk at $R=20$~kpc requires us to constrain the sky brightness to approximately 0.01\%.
Since M31, with a $190\arcmin \times 60\arcmin$ optical disk, is much larger than the WIRCam fields of view, monitoring of the sky zeropoint is only possible by periodically pointing the telescope away from M31, towards blank sky, through \emph{sky-target} (ST) nodding. 
The fundamental compromise of ST nodding observation programs is to balance the cadence of sky sampling with the efficiency of observing the target itself.
Although studies such as \cite{Vaduvescu:2004}, and references therein, provide good guidelines for NIR sky behaviour, no program has attempted to construct a near-IR surface brightness mosaic covering an area as large and faint as M31's disk.

In this program, we have the opportunity to experiment with different ST nodding strategies since observations were taken over the 2007B and 2009B semesters.
An objective of this study is to determine how observational design can improve the construction of a wide-field NIR mosaic by comparing the performance of these two observing strategies.

\begin{figure}[t]
\centering
\includegraphics[width=\columnwidth]{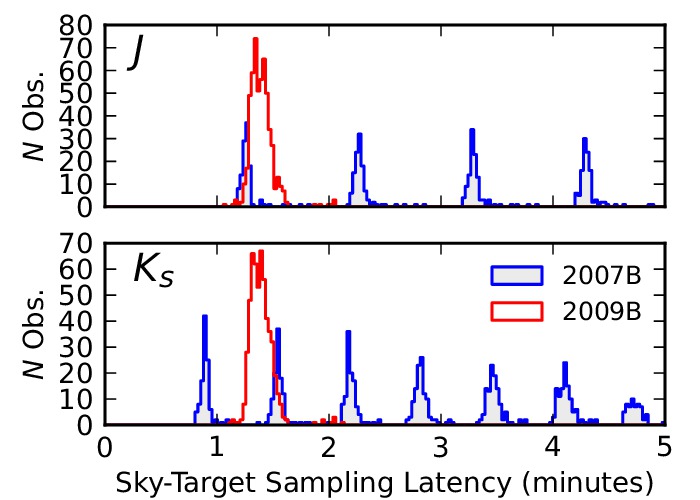}
\caption{Time latency between target observations and sky field sampling in the 2007B and 2009B WIRCam observing runs.
The 2009B program was designed to ensure that no disk sample would be removed by more than 1.5 minutes from a sky sample by using a STTS nodding pattern.}
\label{fig:sky_target_lag}
\end{figure}

\begin{figure}[t]
\centering
\includegraphics[width=\columnwidth]{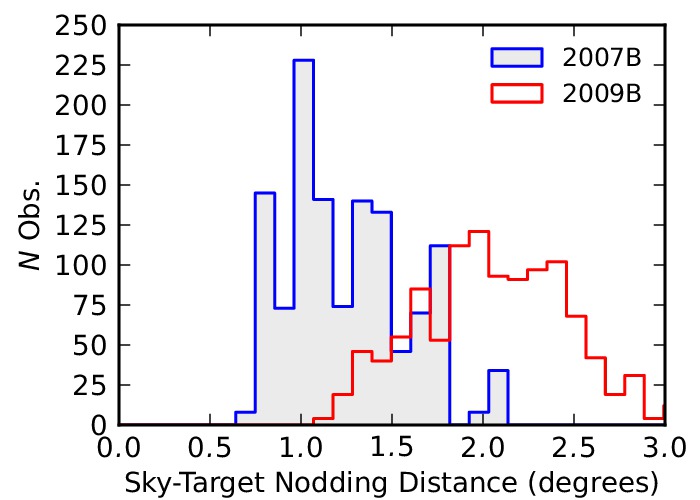}
\caption{Distance between sky and target observations in the 2007B and 2009B WIRCam observing runs.
The larger nodding distance of 2009B is a consequence of sky ring sampling.
The maximum nodding distance across the sky ring was purposely set to $\sim 3$\arcdeg\ to avoid excessive time overheads (see \Fig{fieldmap}).
As such, a given disk field only samples roughly half of the full sky ring.}
\label{fig:sky_target_dist}
\end{figure}

\begin{figure}[t]
\centering
\includegraphics[width=\columnwidth]{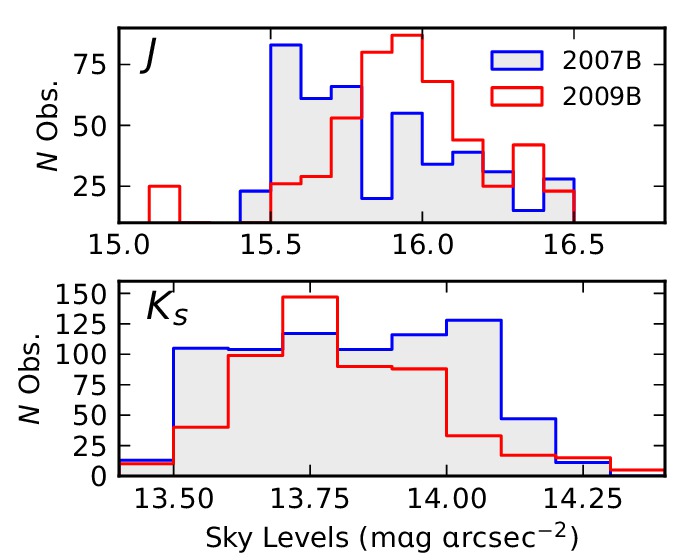}
\caption{Sky levels observed in the 2007B and 2009B programs, as applied to each target field.}
\label{fig:net_sky_level}
\end{figure}

\subsection{2007B Semester}
\label{sec:obs7}

The initial survey was carried out in the 2007B semester by the CFHT Queue Service Observing under photometric conditions.
Here M31 is covered with 27 contiguous WIRCam fields out to the optical radius where $\mu_V=23$ mag arcsec$^{-2}$ at $R=20$~kpc.
The fields are arranged with at least 1\arcmin\ overlap in declination, and approximately 5\arcmin\ overlap in right ascension.
The field configuration is shown in \Fig{fieldmap}.

As shown in Table~\ref{tab:obssummary}, each field was integrated for $16\times 47~\mathrm{s} = 12.5$ minutes in $J$ and $26\times 25~\mathrm{s} = 10.8$ minutes in $K_s$.
These integrations are sufficiently deep for resolved stellar photometry to reach at least 1~mag below the tip of the red giant branch, a crucial requirement for decomposing the contributions of red giant and asymptotic giant branch stars to the NIR light.

The 2007B ST nodding strategy was motivated by a canonical understanding of NIR sky behaviour, since ST nodding sky subtraction had never been attempted on this scale before.
The NIR sky intensity can be expected to change by 5\% in 10~minutes \citep{Adams:1996,Vaduvescu:2004}.
Since the sky itself is $\sim10^3\times$ brighter than the outer disk of M31 in the NIR, a 5\% uncertainty in the background would be fatal to our objective of recovering M31's extended NIR surface brightness.
To constrain sky brightness to within 1\%, we ensured that a sky sample would be no more than 5 minutes removed from a M31 target image.
Given the respective exposure times (chosen so as not to saturate with the sky brightness), this implied a sky ($S$)--target ($T$) observing sequence of $S^3T^8S^3$ in $J$ and $S^5T^{13}S^5$ in $K_s$.\footnote{Superscripts here denote the number of times an observation is repeated in sequence for a given target disk field.}
Four sky fields were chosen (\Fig{fieldmap}), and each disk field was associated with a single sky field.

\subsection{2009B Semester}
\label{sub:obs9}

Analysis of the 2007B data revealed that the adopted sky-target nodding strategy was not sufficient for recovering the M31 surface brightnesses due to uncertainties in the sky background.
Repeatedly sampling one of only four sky fields also proved not ideal.
This motivated a 2009B observing campaign informed by our experiences.

Rather than replicate the 28-field footprint of the 2007B campaign, we observed 12 new fields (see \Fig{fieldmap}) that overlap each other and all of the 2007B footprints, to form a network of well sky subtracted fields.
Thus the 2009B observations augment and calibrate the 2007B NIR mapping.

To improve sky subtraction fidelity, we recognized challenges not fully appreciated in the 2007B survey design.
Not only does the sky background change rapidly in time, it possess a significant spatial structure on the scale of WIRCam fields and larger.
This has two ramifications: the sky level sampled at a sky field \emph{will not} necessarily reflect the sky background present at the disk, and that the sky background in each WIRCam frame has a 2D shape, not simply a scalar level.

This resulted in three principle changes to observing strategy. First, we chose to minimize latency between sky and target observations with a ST$^2$S pattern. That is, each target observation was directly paired with a sky observation taken within 1.5 minutes (\Fig{sky_target_lag}).

Second, we also increased the number of repetitions on each field, so that each field is observed 40 times in each band in a [ST$^2$]$^{20}$S pattern. This repetition enables averaging over spatial sky background structures on the scale of WIRCam fields.

Finally, we employ a pseudo-randomized sky-targeting nodding pattern where no sky field is used repeatedly for a disk field.
In order to maintain rapid telescope nods, only northern sky fields serviced the northern disk, and similarly for the southern fields; the maximum offset on the sky was 3\arcdeg\ (see \Fig{sky_target_dist}).
This non-repetitive sampling of sky fields yielded two possible advantages: 1) when a median sky image is constructed, many \emph{sky shapes} are combined, possibly yielding an intrinsically flatter image of sky (see \Sec{mediansky}), and 2) if there is a coherent structure in the NIR sky, sampling sky fields degrees apart in rapid succession should average out these systematic biases in estimating the sky level \emph{on the disk}.

\section{Image Preparation}
\label{sec:reduction}

The crux of this paper is finding how wide-field NIR mosaics can best be made with the WIRCam instrument on CFHT, and determining what limits the accuracy of our surface brightness maps.
To begin, we note that WIRCam data are offered by CFHT in three progressive stages of \emph{preprocessing} by their \iiwione\ pipeline to allow programmes, such as this one, to re-implement calibration recipes for potentially higher surface brightness accuracy.
The data flavours are: a raw image that is essentially untouched after leaving the instrument (\texttt{*o.fits}); an image that has been corrected for nonlinearity, dark subtracted and flat fielded (\texttt{*s.fits}); and an image that has been sky subtracted, in addition to all the previous treatments (\texttt{*p.fits}).

As sky subtraction is the highest source of error in our program, the middle data product, \texttt{*s.fits}, would appear most amenable as a starting point.
Nonetheless, two \iiwione\ processing stages included in \texttt{*s.fits} products must be handled carefully.

\paragraph{Cross-talk correction} WIRCam integrations prior to March 2008 (includes the 2007B data set, not the 2009B data) suffered from electronic cross talk within the detector.
This cross talk is manifested in repeating rings above and below saturated stars.\footnote{See \url{http://cfht.hawaii.edu/Instruments/Imaging/WIRCam/WIRCamCrosstalks.html}.}
By default, the \iiwione\ pipeline removes this cross talk by subtracting a median of the 32 amplifier slices.
Unfortunately, this algorithm fails in cases where the background has a surface brightness gradient (such as on the disk of M31) and produces an brightness gradient that is stronger than the galaxy surface brightness itself.
Loic Albert (then at CFHT) kindly re-processed our 2007B data set with the cross-talk correction turned off.

\begin{figure}[t]
\centering
\includegraphics[width=\columnwidth]{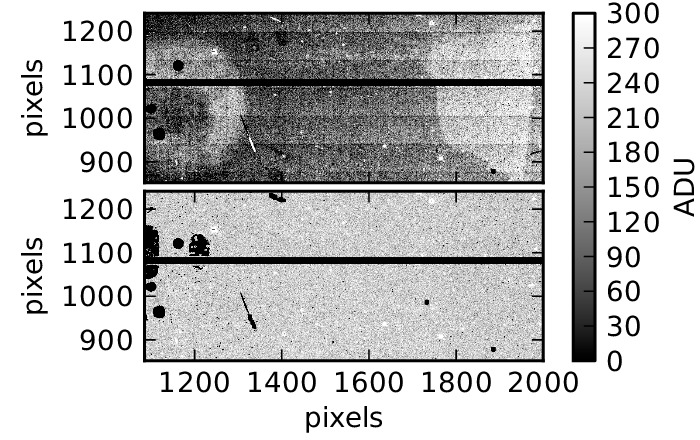}
\caption{Comparison of a WIRCam frame cutout processed with dome flats by the \iiwione\ pipeline (top), and with sky flats (bottom).
Both images are shown in linear counts with identical level ranges.
No median sky subtraction has been applied.
Dome flats leave WIRCam images with dust artifacts (left) and detector surface defects (right). Furthermore, the 64-pixel high horizontal amplifier bands are clearly visible.
Simply using sky flats eliminates these artifacts.
}
\label{fig:flattenedimage_comparison}
\end{figure}

\begin{figure}[t]
\centering
\includegraphics[width=3.5in]{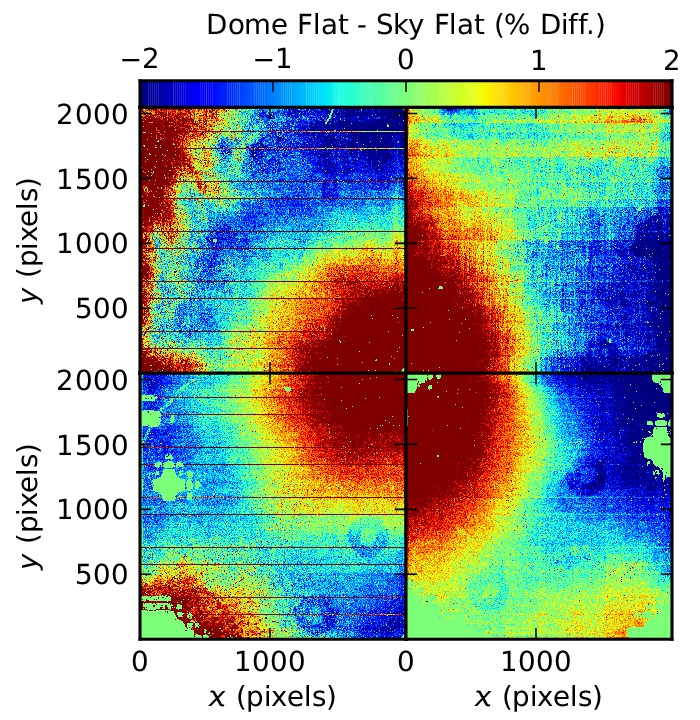}
\caption{Difference image between the 09BQ01 $K_s$ (queue run) sky flat and the \texttt{domeflat\_8302B\_20090728HST143302\_Ks} dome flat used in the \iiwione\ pipeline, in percent.
Using the median night sky illumination over a WIRCam queue run, rather than a dome lamp, results not only in $>2$\% edge-to-edge difference in the large scale WIRCam illumination function, but also in different characterizations of gain structure in the WIRCam detector amplifiers (the 32 horizontal bands in each of the four WIRCam detectors).}
\label{fig:domeflatratio}
\end{figure}

\paragraph{Flat fielding} We discovered that the dome flat fielding offered by \iiwione\ was only accurate to 2\% of edge-to-edge intensity.
The ineffectiveness of WIRCam dome flat fielding is evident in \texttt{*s.fits} images that show significant detector structure, despite having been flat fielded.
An example of this structure is related to the different gain structures of the 32 amplifiers that service independent horizontal bands across each WIRCam detector (see \Fig{flattenedimage_comparison})---a proper flat field capture such gain structure for proper photometry.
Users of fully-processed \iiwione\ \texttt{*p.fits} images do not directly notice the unsuitability of dome flats since images of the detector gain structure embedded in the NIR bright sky background are \textit{subtracted} from the signal as part of the median sky subtraction step.
Yet since pixel gain changes the signal in proportion to incident flux, subtraction is fundamentally the wrong operation to use.
Our remedy is to adopt night sky flats, which use the median night sky as the illumination reference rather than a dome lamp.
The profound difference between dome or night sky flats is shown in \Fig{domeflatratio}.
Our confidence in night sky flats as the \textit{correct} choice for flat fielding WIRCam lies in their ability to remove both large scale illumination features (see \Sec{skyflatstability}) and pixel-to-pixel gain changes across WIRCam (\eg \Fig{flattenedimage_comparison}).
Production of WIRCam sky flats is subtle as we find the flat field structure to be time dependent on sub-hour scales, and the NIR night sky itself is not flat either.
A comprehensive discussion of WIRCam flat fielding is presented in \Sec{skyflatstability}.

\vspace{1em}

Our \androids\ pipeline thus begins with \texttt{*s.fits} data that have been uncorrected for dome flat fielding.
That is, we multiply the \texttt{*s.fits} image with its associated dome flat.\footnote{Dome and twilight flats are made available by CFHT, \url{http://limu.cfht.hawaii.edu:80/detrend/wircam/}.}
The result is an image that retains \iiwione's prescription for dark subtraction, bad pixel masking and non-linearity correction.
The \androids\ WIRCam pipeline then follows the steps described below.

\subsection{Reduction outline}
\label{sec:reduction_outline}

Our ANDROIDS/WIRCam reductions begin by unifying the World Coordinate System across the dataset with \sw{SCAMP} \citep{Bertin:2006}.
Scamp matches stars in Source Extractor \citep{Bertin:1996} catalogs of each WIRCam \texttt{*p.fits} frame both internally (to $\sigma_\mathrm{int}=0\farcs 10$), and against the 2MASS Point Source Catalog \citep{Skrutskie:2006} to a precision of $\sigma_\mathrm{ref}=0 \farcs 15$.
By processing all 4286 frames in the \androids /WIRCam survey simultaneously, \sw{SCAMP} allows an accurate and internally consistent coordinate frame for our mosaic.
\sw{SCAMP} handles this data volume gracefully provided we cull the input star catalogs for stars with $S/N > 100$, and by using the \texttt{SAME\_CRVAL} astrometry assumption that the WIRCam focal plane geometry is stable.

The next steps are flat fielding and median sky subtraction.
For flat fielding, we apply flat field frames built from sky images (sky flats), as opposed to the dome flats and twilight sky flats employed in the \iiwione\ pipeline.
As we explore thoroughly in Section~\ref{sec:flats} and again in Section~\ref{sec:skyflatstability}, sky flats are crucial for tracking flat field evolution in the WIRCam detector.
Next, our pipeline performs preliminary sky subtraction using median sky images, as described in Section~\ref{sec:mediansky}.
Since we apply a new flat field recipe, our pipeline also re-estimates photometric zeropoints by bootstrapping directly against the 2MASS Point Source Catalog; this operation is described in Section~\ref{sec:photocal}.

Finally, while the data have formally been flat fielded and photometrically calibrated, the sky level estimation provided by the sky-target nodding observing scheme is not accurate enough to build a seamless mosaic.
Instead we model sky offsets that enforce internal consistency in the sky levels of target frames.
In Section~\ref{sec:scalar} we discuss the methods for modelling these sky offsets and provide an analysis of their amplitudes in  Section~\ref{sec:scalaranalysis}.

\section{Sky Flat Fielding}
\label{sec:flats}

Dome flats fail to properly calibrate WIRCam data, as evidenced by Figs.~\ref{fig:flattenedimage_comparison}~and~\ref{fig:domeflatratio}.
Sky flats are an appropriate alternative, both because of the abundant sky background (any NIR imaging program can use its own images to build sky flats), and because sky flats more aptly trace detector illumination and gain structure.
The former because skylight traces the same optical path as astronomical sources; the latter because WIRCam's gain structure appears variable.
Sky flats allow detector gain mapping in real-time.
We return to the variability of WIRCam flat fields in Section~\ref{sec:skyflatstability}.

\subsection{Sky Flat Designs}
\label{sec:flatdesign}

Producing sky flats is as simple as median-combining integrations of blank sky. 
The \androids\ sky-target nodding observing strategy provides an abundance of `sky' images for this purpose.
A fundamental decision is the definition of the window of sky integrations that are combined into a sky flat.
Here we present two sky flat designs: labelled \texttt{QRUN} and \texttt{FW100K}.

The first choice assumes that flat fields are stable over a queue run (a continuous installation period of WIRCam).
In this case, all sky integrations taken during a queue run, and through a given filter, are combined into a \texttt{QRUN} sky flat.
This choice is reasonable since dust and optical geometry should be stable during a continuous mounting period.
Further, choosing a large pool of sky images ensures high \sn\, and helps to marginalize over the shape of the sky background.
For the \androids\ program, \texttt{QRUN} skyflats are built from 25--637 sky integrations over several nights (see \Tab{qrunflattable}).

\begin{table}[t]
\centering
\caption{Properties of \texttt{QRUN} sky flats constructed for each CFHT/WIRCam Queue Run (\texttt{QRUNID}).}
\label{tab:qrunflattable}

\begin{tabular}{cccc}
\hline
Filter & QRUNID & $N$ images & $N$ nights \\
\hline
$J$ & 07BQ01 & 24 & 6 \\
$J$ & 07BQ03 & 141 & 9 \\
$J$ & 07BQ05 & 43 & 4 \\
$J$ & 09BQ01 & 111 & 10 \\
$J$ & 09BQ03 & 55 & 3 \\
$J$ & 09BQ08 & 396 & 10 \\
\hline
$K_s$ & 07BQ01 & 35 & 2 \\
$K_s$ & 07BQ03 & 25 & 1 \\
$K_s$ & 07BQ05 & 156 & 6 \\
$K_s$ & 07BQ07 & 133 & 3 \\
$K_s$ & 09BQ01 & 113 & 10 \\
$K_s$ & 09BQ03 & 55 & 3 \\
$K_s$ & 09BQ08 & 637 & 8 \\
\hline
\end{tabular}
\end{table}

\begin{figure}[t]
\centering
\includegraphics[width=\columnwidth]{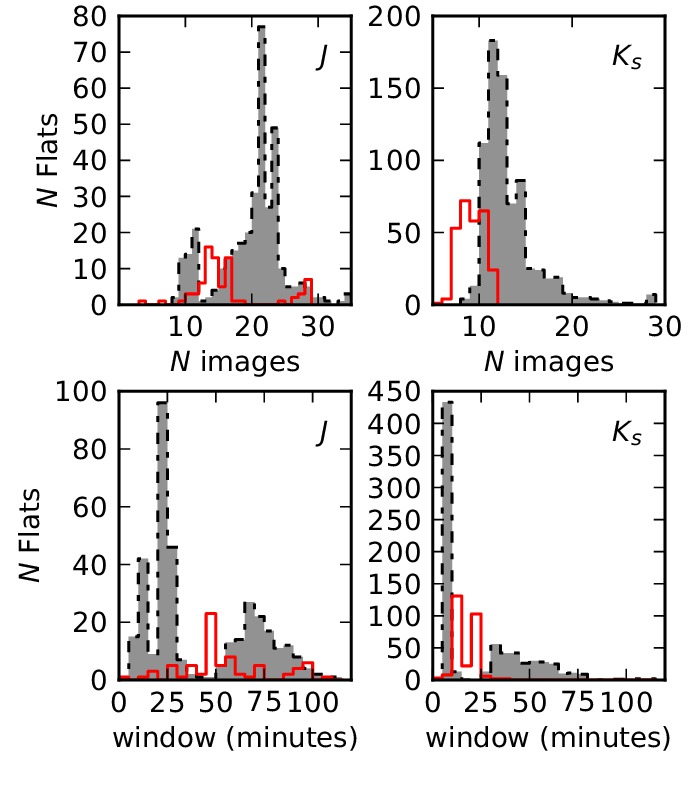}
\caption{Characteristics of \texttt{FW100K} real-time sky flats, meeting criteria of at least 100,000 sky ADU, and spans of less than two hours.
Distributions of 07B sky flats plotted as red outlined histograms, 09B sky flats are plotted as shaded histograms.
The number of sky images required to build a sky flat (top), and the time span from first to last sky image depends on sky brightness and the sky-target nodding pattern (bottom).
Real-time sky flats can consequently be refreshed in as little as 10 minutes, or span the order of an hour.}
\label{fig:fw100k_summary}
\end{figure}

An alternative design choice assumes that WIRCam's illumination function and detector gain structure is unstable even over short periods.
Here the objective is to make many sky flats that reflect the real-time flat field function---we call these Real-Time Sky Flats.
Our first real-time sky flat, labelled \texttt{FW100K}, is designed such that the pool of sky images reaches cumulative sky levels of at least 100,000 ADU, or that the time span from first to last sky integration be no longer than two hours.
Properties of these sky flats---the number of images required to build them, and the width of the windows they cover---are shown in \Fig{fw100k_summary}.
Given the 07B $J$-band ST nodding pattern, 15 sky integrations are accumulated in 50 minute windows, whereas the more frequent nodding in the 09B campaign shortened this window to 20 minutes (though as long as 50--90 minutes in dark sky conditions).
The brighter $K_s$ sky calls for just 7--13 integrations in 07B, or 10--20 integrations in the 09B campaign.
This number of $K_s$ sky samples was accumulated within 10--30 minutes in 07B, or 10--70 minutes in 09B.

\subsection{Building Sky Flats}
\label{sec:flatbuilding}

Given the ensemble of sky integrations, our next task is to scale the intensity of each image.
This scaling obeys three requirements: 1) each image frame in the median stack is at the same level, 2) each WIRCam detector has a unified zeropoint, and 3) the sky flat across the whole array is flux normalized.\footnote{It is also acceptable to establish chip-to-chip zeropoint offsets using differential 2MASS photometry, rather than from sky surface brightness. In \Sec{detector_zp} we establish the equivalence of the two methods.}
This scaling is determined by the median pixel level measured on each detector for each sky integration---let us denote these median levels as $\alpha_{i,j}$ for the $i$th sky image's level in detector $j$, $j=1, 2, 3, 4$.
To avoid bias in the sky estimate, we mask any pixels that do not sample blank sky.
\sw{Source Extractor} \citep{Bertin:1996} is used to define stars and background galaxies (we use \iiwione\ \texttt{*p.fits} images to detect and mask sources), while hand-drawn polygon regions cover diffraction spikes and the diffuse halos around bright galactic stars. These masks, along with the \iiwione\ bad pixel mask, are combined with \sw{WeightWatcher} \citep{Marmo:2008}.

\begin{figure}[t]
\centering
\includegraphics[width=\columnwidth]{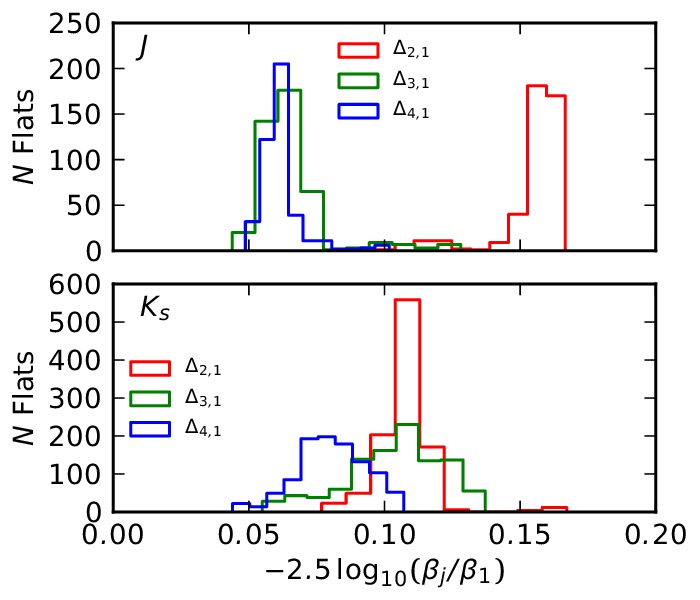}
\caption{Distribution of real-time sky flat scaling factors, measuring detector-to-detector zeropoint differences relative to detector \#1 (red: \#2, green: \#3, blue: \#4) in $J$ and $K_s$ bands.
}
\label{fig:fw100k_zpdiff}
\end{figure}

From the ensemble of images produced by an individual WIRCam detector, we compute the median sky level: $\beta_j = \mathrm{median}(\alpha_{1,j}, \alpha_{2,j}\ldots \alpha_{n,j} )$.
Further, we also compute $S$, the median of all median detector levels: $S=\mathrm{median}(\beta_j)$.
Then each the sky image is scaled by the factor $f_{ij} = \mathrm{median}(\beta_1, \beta_2, \beta_3, \beta_4) / (\alpha_{ij} S)$.
Note that the factor $\alpha_{ij}^{-1}$ normalizes each image to the same level for stacking, while the ratio $\beta_j / S$ adjusts the level of each detector according to detector-to-detector zeropoint offsets.
Indeed, detector-to-detector zeropoint offsets can be measured as $-2.5 \log_{10}(\beta_i / \beta_1)$, as shown in \Fig{fw100k_zpdiff}.
WIRCam detector \#1 (North-West quadrant) is clearly the most sensitive, while the dispersion indicates the level of gain variability in WIRCam.
We also note systematic differences in spectral response of WIRCam detectors in the $J$ and $K_s$ of order 5\%.

The flat itself is built by median combination.
Median combination of a stack of hundreds of $2048\times2048$ pixel images, each with a weightmap masking astronomical sources, is computationally intensive.
A convenient solution is to use \sw{Swarp} \citep[an image-mosaicing software package,][]{Bertin:2002} in a mode that combines images pixel-to-pixel.

\section{Median sky image subtraction}
\label{sec:mediansky}

Since M31 is much larger than individual WIRCam fields, sky background is subtracted (to first order) using the sky levels found in contemporary sky images.
Section~\ref{sec:Observations} described the sky-target nodding sequences chosen for the 2007B and 2009B observing campaigns. 
Although a scalar sky level can be estimated from a sky image, and subtracted from the paired target images, it is common to construct a median sky image, the same size as the WIRCam frames, and subtract this 2D image from target images.

Independent median sky images for each WIRCam detector are produced by choosing a sky image (the primary sky image) and four other sky images taken at adjacent times.
Across each image, the median sky intensity is recorded.
A Source Extractor object mask, as used in \Sec{flatbuilding} for flat fielding, removes bias from astrophysical sources.
Each sky image is \emph{additively} scaled to a common intensity level, allowing differences in overall sky amplitude to be ignored by the median combination.
As described in \Sec{flatbuilding}, \sw{Swarp} is used to median-combine the sky images, taking into account non-sky pixel masks.
Since these median sky images record only low-frequency spatial information, these median sky images are smoothed with a Gaussian kernel (note this is quite different from the function of median sky images applied to dome-flat processed WIRCam data, where median sky subtraction also removed pixel-to-pixel artifacts).
This median sky image is then additively scaled back to the original level of the primary sky image.

Each science image is sky subtracted by first identifying the median sky frame whose primary image was taken most closely in time.
That paired median sky frame is subtracted from the image.

In \Sec{skyflatstability} we follow up on this operational discussion to consider the shapes of median sky images. In particular, we connect the role of median sky frames to the stability of different recipes of sky flats described in \Sec{flats}.

\section{Photometric calibration}
\label{sec:photocal}

Since the sky flats used by this \androids/WIRCam data set differ from the dome flats employed by the \iiwione\ pipeline by 2\% in edge-to-edge level (see \Fig{domeflatratio}), new photometric zeropoints must be established.
The Two Micron All Sky Survey (2MASS) Point Source Catalog (PSC) \citep{Skrutskie:2006} provides a convenient photometric system to bootstrap against.
WIRCam pointings ($20\arcmin \times 20\arcmin$) in this survey contain typically $\sim 500$ 2MASS stars.
Although these are not standards, the ensemble of 2MASS stars may be treated as such.
Since the disk of M31 is crowded, and 2MASS has poor resolution ($1\arcsec$~per pixel), we consider 2MASS point source measurements to be unreliable on the disk.
Instead, all zeropoints are measured against sky images, and those calibrations are applied to paired disk images (analogous to the median sky subtraction procedure, described in Section \ref{sec:mediansky}).

Photometry of 2MASS stars in the uncrowded sky fields is obtained with Source Extractor \citep{Bertin:1996}.
We use the \texttt{AUTO} photometry mode to capture the full stellar light without using aperture corrections.
Objects in the 2MASS PSC are matched to Source Extractor detections by position using J. Sick's \sw{Mo'Astro}\footnote{\url{http://moastro.jonathansick.ca}} Python package, that manages the full 2MASS PSC in a MongoDB database.
The 2MASS PSC contains many galaxies, and many 2MASS sources are saturated in our deeper WIRCam images.
Thus we select stars with $m_J < 14$ or $m_{K_s} < 15$ magnitudes, and stars with FWHM $<1$\arcsec.
Additionally, we select sources with $J-K_s < 0.8$ (typical of foreground Milky Way stars) as we observe larger zeropoint residuals in redder stars.
After filtering, typically 200 matched 2MASS sources remain in typical WIRCam images.

To fit a photometric zeropoint, we bootstrap directly against 2MASS photometry visible in sky images.
This practice obviates any airmass dependence, since each fitted zeropoint, $m_0$ is instantaneous.
Given sample of matching 2MASS and instrumental photometry, the instrumental zeropoint is estimated as median:

\begin{equation}
  \label{eq:photcal}
  m_0 = \langle m_\mathrm{2MASS} + 2.5 \log_{10}(\mathrm{ADU}/T_\mathrm{exp}) - A (J-K_s)_\mathrm{2MASS} \rangle.
\end{equation}

The $A$ term permits a linear colour transformation between the 2MASS and WIRCam bandpasses.
Although we could fit a colour transformation coefficient $A$ for each image, the short $J-K_s$ colour baseline makes such measurements unreliable.
Instead we adopt $A_J = 0.05$ and $A_{K_s} = -0.05$ (K. Thanjuvar, priv.\ comm.) based on modelling stellar spectra with the 2MASS and CFHT/WIRCam transmission functions.
For typical M31 RGB stars with $J-K_s\sim 1$, this colour transformation is a 0.1~mag effect.

Given that 2MASS stars in each image have photometric uncertainties $0.05 \lesssim \sigma_{\mathrm{2MASS~mag}} \lesssim 0.3$, the typical statistical zeropoint uncertainty, $\sigma_{m_0}$, is 0.1~mag in a single image.
Since $m_0$ is fit for sky images, zeropoints for science images are interpolated as the median of a sliding window of sky images large enough for the statistical uncertainty of the median $m_0$ to be reduced below 0.01 mag.

\subsection{Detector-to-Detector zeropoint consistency}
\label{sec:detector_zp}

\begin{figure}[t]
\centering
\includegraphics[width=\columnwidth]{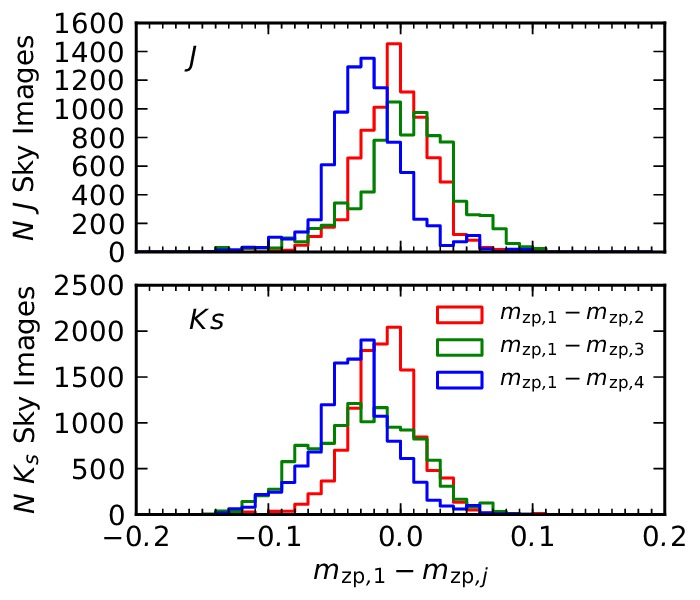}
\caption{Distribution of mean detector-to-detetector zeropoint offsets for sky images processed by real-time sky flats. Zeropoint offsets between detectors \#1--\#2, --\#3, and --\#4 are plotted as
red, green and blue histograms, respectively, for the $J$-band (top) and $K_s$-band (bottom).}
\label{fig:fw100k_chip_to_chip}
\end{figure}

Our sky flats are designed to unify the zeropoints of the four WIRCam detectors by scaling according to the modal sky levels seen on each detector (see~\Sec{flats}).
The accuracy of this calibration can be verified by comparing photometric zeropoint estimates for individual WIRCam detectors.
Figure~\ref{fig:fw100k_chip_to_chip} shows the distribution of mean detector-to-detector zeropoints offsets observed in images processed by real-time sky flats. We find that zeropoints are consistent within $\pm 0.1$ mag, although we detect a possible systematic bias between detectors \#1 and \#4 at the level of $0.03$~mag.

\section{Sky Offset Optimization}
\label{sec:scalar}

\begin{figure}[t]
\centering
\includegraphics[width=3.5in]{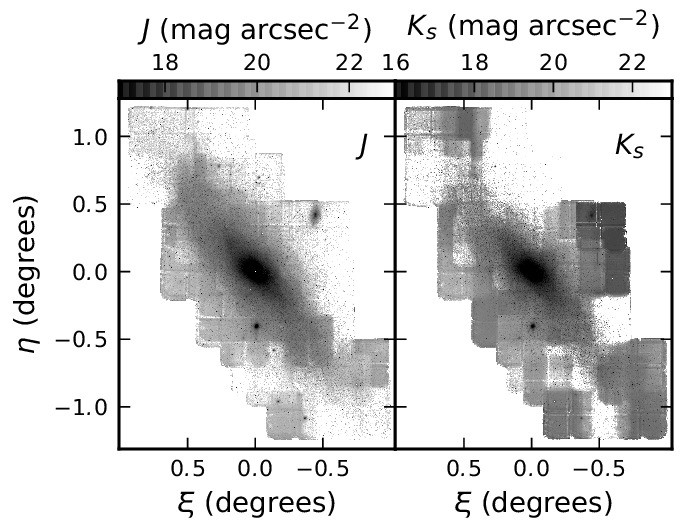}
\caption{Median sky subtracted WIRCam $J$ (left) and $K_s$ (right) mosaics of M31.}
\label{fig:raw_mosaics}
\end{figure}

Despite our attention thus far to real-time sky flat fielding, and median sky image subtraction, we have yet to recover the true NIR surface brightness of M31.
In \Fig{raw_mosaics}, we plot mosaics (assembled using \sw{Swarp}) from \androids\ frames processed with the pipeline discussed in \Sec{reduction}--\Sec{photocal}.
Although these basic image preparations dramatically improve image quality compared to the \iiwione\ pipeline (particularly with regards to flat fielding performance), classical NIR sky subtraction still fundamentally limits accuracy.
Classically, the true value of the sky on M31's disk is lost by the temporal and spatial variations of skyglow between disk and sky field observations.

We now demonstrate how the residual sky bias in each observation can be inferred from information in the overlaps of image pairs in the mosaic.
Each classically sky subtracted image of the M31 disk is a combination of the true surface intensity, $I_i$, and a residual sky intensity, $\epsilon_i$.
Consider a pair of images, $i$ and $j$, that overlaps the galaxy: their difference is $(I_i+\epsilon_i) - (I_j+\epsilon_j) = \epsilon_i - \epsilon_j$.
Given this measurement $\epsilon_i - \epsilon_j$ of residual sky intensity, we introduce \emph{sky offsets}, $\Delta$, for each observation so that

\begin{equation}
    (I_i + \epsilon_i - \Delta_i) - (I_j + \epsilon_j - \Delta_i) \rightarrow 0,
\end{equation}

\noindent and the intrinsic intensities cancel, $I_i - I_j = 0$.
Given a single pair of images, the inference of $\Delta_i$ and $\Delta_j$ is degenerate given the single difference image, $\epsilon_i-\epsilon_j$.
In a mosaic, each image is coupled with many other images, and the mosaic itself can be considered as a network of coupled images.
Thus by optimizing a set of sky offsets for all images simultaneously that minimizes all image-to-image differences in a mosaic, a robust of set of $\Delta_i$ can be established.

\subsection{Implementations}
\label{sec:offset_algos}

The sky offset optimization summarized above is difficult to implement because it is over-constrained, requiring a non-linear optimization.
Any single image shares a network with others, and any error in the surface brightness \emph{shapes} of fields will result in an imperfect match.
In this work, we consider two implementations of sky offset optimization.
First, we review the \sw{Montage} software package in \Sec{montage_algo}, and then introduce an alternative in \Sec{msrnm_algo}.

\subsubsection{Montage}
\label{sec:montage_algo}

\sw{Montage} is a FITS mosaicing package \citep{Berriman:2008} originally written for the 2MASS survey that includes sky offset estimation (background rectification, in their terminology) functionality.
Images are rectified onto a mosaic pixel grid, and difference images are computed among overlapping images.
Sky offsets are then chosen iteratively by looping through each image pair and choosing the offset needed to minimize the difference image of that pair, counting previous sky offset estimates.
That is, at each step the sky levels of the two images are increased and decreased by half the total intensity difference.
Sky offsets are refined over several loops through the entire set of overlapping image pairs until convergence is reached (that is, once incremental adjustments to sky offsets diminish below user-specified threshold).
Although this iterative implementation of sky offset optimization is elegant, its accuracy has never been formally analyzed in literature, to our knowledge.
In particular, we are interested in the robustness of Montage sky offsets against local minima in the $N$-dimensional solution space of sky offsets, given a mosaic of $N$ independent images.

\subsubsection{A New Implementation based on the Multi-Start Reconverging Downhill Simplex}
\label{sec:msrnm_algo}

As an alternative to the iterative algorithm used by \sw{Montage}, we introduce here an implementation based on the Downhill Simplex algorithm \cite[][hereafter, NM]{Nelder:1965}.
Like \sw{Montage}, our pipeline begins with a set of images resampled to a common pixel grid in an Aitoff projection with the native WIRCam pixel scale (0\farcs 3 per pixel).
For this we use \sw{Swarp} in resample-only mode.
We identify overlaps between images in a brute-force fashion according to their frames in the mosaic pixel space, defined by the \texttt{CRPIX}, \texttt{NAXIS1} and \texttt{NAXIS2} header values of the resampled images.

For each overlapping image pair, we compute a difference image, and ultimately a median difference, $\langle \vect{I}_i - \vect{I}_j \rangle$.
While computing the median difference, we mask bad pixels using weight maps (propagated by \sw{Swarp}) and expand this mask with sigma clipping.
Along with a difference estimate, we also record the area $A_{ij}$ of unmasked pixels in the overlap, and the standard deviation of the difference, $\sigma_{ij}$.

Let us define the objective function that encapsulates the effect of scalar sky offsets on reducing the intensity difference between coupled images:

\begin{equation}
    \mathcal{F} \left(\Delta_1,\ldots,\Delta_n \right) = \sum_{i,j} \mathcal{W}_{ij} \left( \langle \vect{I}_i - \vect{I}_j \rangle - \Delta_i + \Delta_j \right)^2,
    \label{eq:objf}
\end{equation}

\noindent which we intend to minimize by finding the optimal set of scalar sky offsets $\Delta_i$ for each of the detector fields $i$.
Note that each coupled image pair is its own term in the objective summation, and that there are as many degrees of freedom ($\Delta_i$) as there are images in the mosaic.
Each coupling is tempered by a weighting term $\mathcal{W}_{ij}$:

\begin{equation}
    \mathcal{W}_{ij} = \frac{A_{ij}}{\sigma_{ij}},
\end{equation}

\noindent so that more priority is given to couplings of larger areas ($A_{ij}$), and small standard deviations of their difference images ($\sigma_{ij}$).

Note that the objective function in \Eq{objf} puts no constraint on the net sky offset: $\sum \Delta_i$.
Assuming that sky subtraction errors are normally distributed, and not biased, sky subtraction offsets should not add a net amount of flux to the mosaic.
Fortunately, it is possible to impose this constraint \textit{post facto} by subtracting the mean offset from the sky offsets:

\begin{equation}
    \Delta_i^* = \Delta_i - n^{-1}\sum_{j=1}^n \Delta_j.
    \label{eq:netzero}
\end{equation}

\noindent In the limit that sky offsets $\Delta_i$ are drawn from a Gaussian distribution, with standard deviation $\sigma_\Delta$, the absolute brightness of the whole mosaic will be uncertain by $\sigma_\Delta / \sqrt{N_\mathrm{images}}$.
The consequences of this uncertainty are revisited in \Sec{systematics}.

Given the image coupling records, we optimize the set of $\Delta_i$ using the NM downhill simplex.
This NM algorithm is naturally multi-dimensional and does not require knowledge of the gradient of the objective function.
Instead, the NM algorithm operates by constructing a geometric simplex of $N+1$ dimensions that samples the sky offset parameter space.
By evaluating the objective function at each vertex of the simplex, the NM algorithm adapts the simplex shape to ultimately contract upon a minimum.

However, NM has two weaknesses.
First, it is a greedy optimizer that will converge into any local minimum, without necessarily seeking the global minimum.
Second, for high-dimension optimizations (many fields in the mosaic), the NM may fail to converge in a reasonable number of objective function evaluations \citep{Neumann:2006}.
Without solving these issues, a minimization of \Eq{objf} with an off-the-shelf NM code yields a mosaic with obvious discontinuities across fields.
To solve the first problem, we develop a Multi-Start Reconverging NM (MSRNM) downhill simplex driver.
This algorithm, outlined below, allows the NM to cumulatively cover a larger portion of parameter space to probabilistically ensure convergence into a global optimum.

\paragraph{Multi-start} To cover a large portion of parameter space, and protect against starting near a false minimum, we start several independent simplex runs from random points in parameter space.
We find that $N_s=50$, and possibly fewer, starts are quite sufficient for an optimization with 39 sky offset parameters (such as the fitting of $\Delta_B$ block offsets in mosaic, see \Sec{hierarchical_algo}).
For each start, an initial simplex is generated randomly.
Since each point in the $N$ by $N+1$ simplex is a suggested sky offset for a given field, each offset is randomly sampled from the expected distribution of image-to-image sky offsets.
That is, a normal distribution of mean zero intensity, and standard deviation of $\sigma_\mathrm{start}$.
To ensure that the parameter space is well covered, we design $\sigma_\mathrm{start}$ to be $3\times$ greater than the dispersion of image-to-image differences.

Each initial simplex is allowed to converge according to the NM algorithm.
A simplex is deemed to have converged once each point has changed by no more than $10^{-6}$ of the previous iteration.

\paragraph{Restart} As suggested in \cite{Press:2007}, we restart the converged simplex to ensure reconvergence and protect against false minima.
Upon each convergence, the optimal point in the simplex, $\vect{p}$, is recorded.
A new simplex is then generated where one vertex is $\vect{p}$, and the rest are $\vect{p}+\vect{\delta}$ where $\vect{\delta}$ is a normal random variable of mean zero, and standard deviation $\sigma_\mathrm{restart}$.
That is, the simplex of the restart retains one vertex upon the previously found minimum, while the other vertices surround that minimum.
If the minimum is indeed the global minimum, then the NM algorithm will quickly reconverge.
More often than not, the other random vertices will probe an even better parameter space, causing the NM restart to converge upon a new minimum.
When a new minimum is found, the process of restarting is repeated until the same minimum is consecutively arrived upon.
Our sky offset optimizations for 39 blocks typically require $\sim1000$ restarts before converging definitively.
Given the large number of restarts, we find that this reconvergence feature is extremely effective at eluding false minima in our optimizations.
We set $\sigma_\mathrm{restart}$ to $2\times$ the dispersion of image-to-image differences.
Like $\sigma_\mathrm{start}$, there is flexibility in choosing $\sigma_\mathrm{restart}$; generally $\sigma_\mathrm{start}$ should be on the order of the observed image-to-image difference dispersion, not necessarily as large as $\sigma_\mathrm{start}$.

Note that each simplex start and series of subsequent restarts can be performed in parallel.
Once all simplex runs are complete, the set of sky offsets belonging to the run that yielded the smallest value of the objective function is adopted.

\subsection{Hierarchical sky offset optimization}
\label{sec:hierarchical_algo}

Sky offset optimization is challenging because of dimensionality.
Each image frame, and the associated scalar sky offset $\Delta_i$, represents an additional dimension in the optimization.
Considering that the WIRCam array produces four image frames for every exposure, the combined 2007B and 2009B data sets consist of 3924 $J$ and 4972 $K_s$ image frames covering the disk of M31.
Such a large optimization is computationally ambitious, but also unnecessary.
Our sky optimization algorithm breaks the optimization of sky offsets into three sequential steps, which we call \emph{hierarchical sky offset optimization}.

The 2007B and 2009B WIRCam surveys include a total of 39 \emph{fields} across the M31 disk (illustrated in \Fig{fieldmap}).
Each WIRCam field is imaged with four detectors, arranged in a $2\times 2$ grid.
Let us define a \emph{detector field} as the collection of images taken with a given detector, at a given field.
Images in a detector field all have the greatly simplifying property of overlapping across a common section of the image frame.

Thus our M31 WIRCam mosaics are assembled by applying sky offsets in three levels of hierarchy.
In the first level we combine the frames in a detector field to produce a \emph{detector field stack}; these offsets are labelled as $\Delta_F$.
The combined 2007B and 2009B surveys have 156 such stacks per filter.
In the second level, the four detector field stacks within each field can be fitted into a \emph{block} using offsets labelled as $\Delta_S$.
A block is a fundamental unit of the mosaic, as all images that are combined within a block were observed under contemporaneous sky conditions.
Finally, in the third level, the 39 blocks can be fitted into a galaxy-wide mosaic for each filter using offsets labelled as $\Delta_B$.
The net scalar sky offset applied to each frame is thus $\Delta_\Sigma = \Delta_F + \Delta_S + \Delta_B$.
In both the second and third levels, we use the MSRNM scheme to optimize the set of $\Delta_S$ and $\Delta_B$ offsets, respectively.
Note that for detector field stacks it is sufficient to simply compute a mean surface brightness across all frames, and directly compute offsets ($\Delta_F$) between between the levels of each frame and the mean level.

\section{Analysis of Scalar Sky Offsets}
\label{sec:scalaranalysis}

\begin{figure}[t]
	\centering
		\includegraphics[width=3.5in]{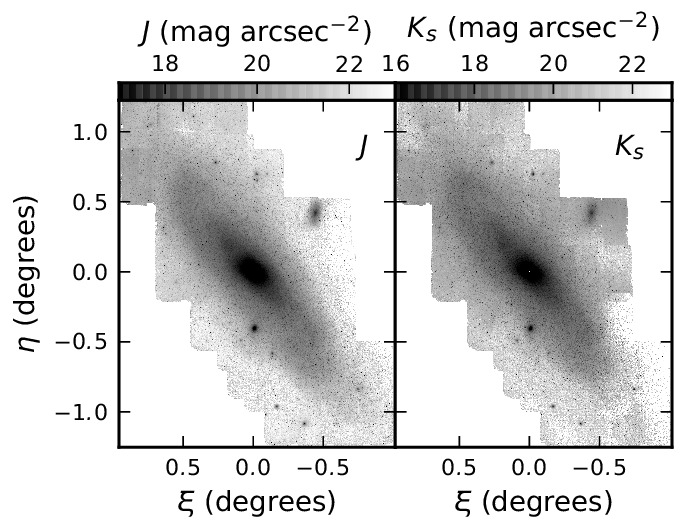}
	\caption{Scalar-sky fitted WIRCam $J$ (left) and $K_s$ (right) mosaics of M31. Note the qualitative improvement compared to the original, median sky-subtracted, images in \Fig{raw_mosaics}.}
	\label{fig:scalar_mosaics}
	\end{figure}

The fruits of our WIRCam pipeline and sky offset optimization are presented in \Fig{scalar_mosaics}.
Compared to our mosaics without sky offsets, \Fig{raw_mosaics}, the sky offset optimization is clearly essential for assembling wide-field NIR mosaics.
Note that these mosaics are not yet perfect; field-to-field discontinuities at a level of 0.05\% of sky remain, and large-scale sky residuals perturb the outer M31 disk.

\subsection{Amplitudes of Sky Offsets}
\label{sec:offset_amplitudes}

The distribution of scalar sky offsets provides an excellent characterization of sky subtraction uncertainties when using sky-target nodding.
Recall that sky offsets are optimized hierarchically: WIRCam frames are fitted to stacks, stacks are fitted into blocks of four contemporaneously-observed WIRCam detector fields, and these blocks are fitted into a mosaic.
\Tab{offset_hierarchy} lists the standard deviations of these offset distributions with respect to the typical sky level observed in the $J$ and $K_s$ bands.

Note that the sky offsets, as a percentage of sky level, are comparable in the $J$ and $K_s$ bands, despite the skyglow being $\sim 4\times$ brighter in $K_s$ than $J$ (see \Fig{net_sky_level}).
This indicates that spatio-temporal variations in the NIR sky are monochromatic.

Within the hierarchy of sky fitting, simply fitting frames to a stack (with $\Delta_F$) is most significant: a correction on the order of 2\% of the sky intensity.
Fitting blocks into a mosaic ($\Delta_B$) is a further $\sim 1$\% correction.
Overall, the temporal and spatial lags of sky-target nodding induce a 2\% uncertainty in the sky level at the target.
It is this level of uncertainty that sky offset optimization must diminish to transform uncorrected mosaics (\Fig{raw_mosaics}) into ones that reproduce the disk with fidelity (\Fig{scalar_mosaics}).

Note that offsets to fit a stack into a block ($\Delta_S$) of four detector field stacks are smallest: 0.1\% of the sky level.
This suggests that on the scale of the $2\times 2$ WIRCam array, the contemporaneously observed detector frames are subjected to nearly identical biases in sky background.
Stack offsets, then, arise from uncertainties in the pipeline's measurement of the sky level from single frames in two stages: estimating detector-to-detector zeropoint offsets from frame sky levels (\Sec{flatbuilding}) and again when subtracting a median sky frame (\Sec{mediansky}).
Indeed, in \Sec{skyflatstability} we show that median sky images have shape amplitudes of 0.3\% of the sky level and that individiual frames have surface brightness shapes that are uncertainty at a level of 0.2\%; $\Delta_S$ sky offsets are thus a consequence of the limited surface brightness flattness across a WIRCam frame.

A comparison between the net sky offsets applied to the 2007B and 2009B data sets is provocative.
Although the 2009B dataset employed rapid sky-target nodding to minimize temporal lags between disk and sky sampling, the magnitude of sky offsets in the 2007B and 2009B semesters is comparable.
This implies a limit to the absolute sky level accuracy that can be expected: the minimal 40~sec lag between sky and target samples, combined with a 1--$2\arcdeg$ nod across the sky, allows the sky level to change by 2\%.
Reducing this latency, and this nodding distance, is impossible in WIRCam observations of M31.
\emph{Thus, by the metric of \Tab{offset_hierarchy}, the expensive 2009B observing approach did not pay off.}

\begin{table}[t]
\centering
\caption[Hierarchy of scalar sky offsets]{Hierarchy of scalar sky offsets (using \texttt{FW100K} RT flat fielding, and median sky subtraction).
The `Total' sky offsets track the net offset of individual WIRCam image frames into the fitted mosaic.
$\langle I_\mathrm{sky}\rangle$ is taken as the instantaneous sky level for the images being sampled (see \Fig{net_sky_level} for the distribution of levels).
Offset distributions are also presented in units of the WIRCam mosaics, DN, corresponding to a zeropoint of 25 mag.}
\label{tab:offset_hierarchy}
\begin{tabular}{ll|rr}
% \hline
&  & $J$ & $K_s$ \\ % \cline{3-4} \cline{5-6}
% \hline
Offset Type & Sem. & $\frac{\sigma_\Delta}{\langle I_\mathrm{sky}\rangle }$ (\%) & $\frac{\sigma_\Delta}{\langle I_\mathrm{sky}\rangle }$ (\%) \\
\hline
\multirow{2}{*}{$\Delta_F$} & 07B & 2.53 & 2.29 \\
% \hline
& 09B  & 1.91 & 1.88 \\
\hline
\multirow{2}{*}{$\Delta_S$} & 07B & 0.11 & 0.05 \\
% \hline
& 09B & 0.11 & 0.06 \\
\hline
\multirow{2}{*}{$\Delta_B$} & 07B & 1.24 & 0.94 \\
% \hline
& 09B & 0.66 & 1.13 \\
% \hline
\hline
\multirow{2}{*}{$\Delta_\Sigma$} & 07B & 2.73 & 2.44 \\
% \hline
& 09B & 1.98 & 1.96 \\
% \hline
\end{tabular}

\end{table}

\subsection{Acceptability of Sky Offsets}
\label{sec:offset_acceptability}

\begin{figure}[t]
\centering
\includegraphics[width=\columnwidth]{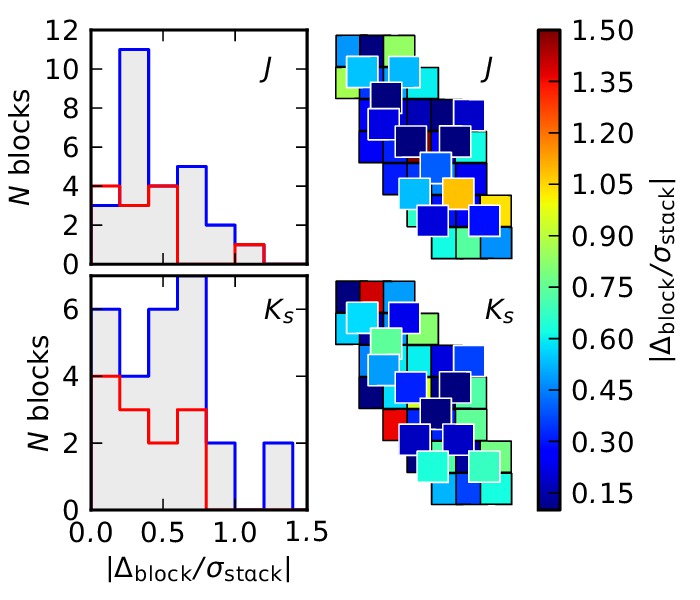}
\caption{Acceptability of $J$ and $K_s$ scalar sky offsets between blocks, as measured by the ratio of $\Delta_B/\sigma_{\Delta_F}$, plotted as histograms and field maps. Shaded blue and red-outlined histograms distinguish blocks observed in 2007B and 2009B, respectively. Scalar sky offsets required for blocks are consistent with the sky level uncertainties of single frames, given sky-target nodding sky subtraction.}
\label{fig:offset_ratio_map}
\end{figure}

Recall that scalar sky offsets were initially introduced as intensity increments to overcome uncertainty in the sky level of detector field stacks.
For sky offsets to be considered acceptable, we demand that the offsets applied to blocks, $\Delta_B$ be consistent with the sky level uncertainty of the blocks themselves.
We can conservatively measure the sky uncertainty as the dispersion of $\Delta_F$ frame offsets in a stack: $\sigma_{\Delta_F}$.
If sky offsets fitted between blocks are statistically permissible, then $\Delta_B \lesssim \sigma_{\Delta_F}$.
In \Fig{offset_ratio_map}, we plot field maps (in the same spatial configuration as \Fig{fieldmap}) painted with the values of $\Delta_B / \sigma_{\Delta_F}$ for each block in the $J$ and $K_s$ mosaics.
The sky offsets are indeed distributed within the uncertainty budgeted by $\sigma_{\Delta_F}$: the sky offsets are statistically acceptable.

\begin{figure*}[t]
\centering
\includegraphics[width=\textwidth]{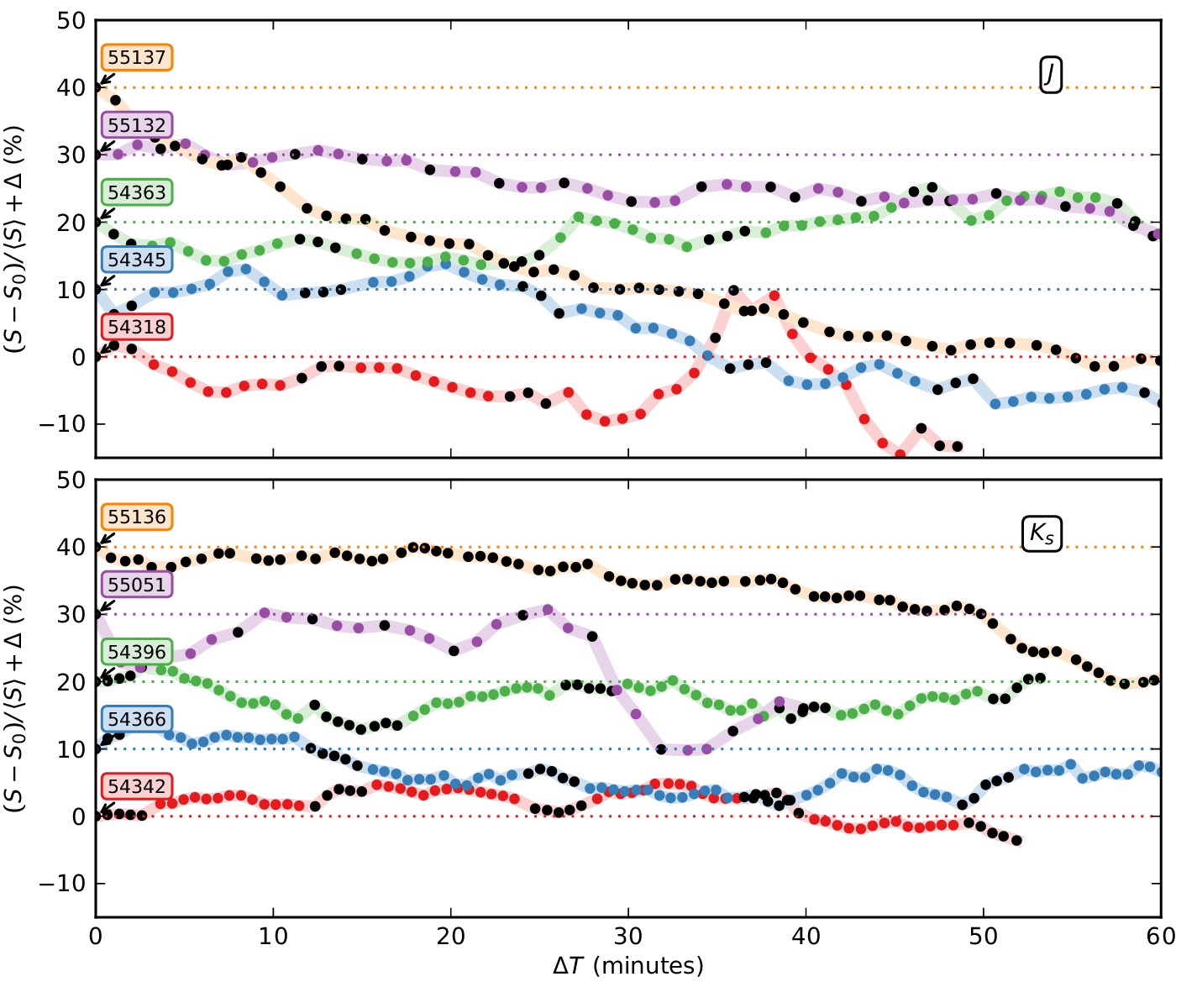}
\caption{Sky level time series for selected nights in the $J$ (top) and $K_s$ bands (bottom) as a percent difference of each night's initial sky level (with a 10\% offset between nights).
Individual nights are tagged by a modified Julian date (MJD).
Directly measured sky levels are plotted in black, while sky levels estimated on the disk (with scalar sky offsets) are coloured dots. The continuity between real and estimated sky levels demonstrates the validity of sky offsets.}
\label{fig:skylevel_timeseries}
\end{figure*}

Another demonstration of the veracity of these sky offsets is given in \Fig{skylevel_timeseries}, where we plot a time series of both directly measured sky levels, and sky levels interpolated on disk observations via sky offsets.
Note the remarkable continuities of the sky level from sky-to-disk observation; compared to the sky level time series of stationary sky fields, the variations in the sky level interpolated on the disk appear real.
Through the sky target nodding and sky offset optimization, we \emph{have} effectively measured the sky level on M31.
The variability of the sky in these time series will be further examined in \Sec{offsetevo}.

\subsection{Residual Image Level Differences}
\label{sec:residual_diffs}

Although scalar sky offsets are statistically valid, they are not perfect prescriptions against the sky subtraction uncertainties of each image stack---that much is visually true.
A measure of the limitation of sky offset fitting are the residual image level differences between coupled blocks, $(\vect{I}_i - \Delta_{B,i}) - (\vect{I}_j - \Delta_{B,j})$, after the sky offsets $\Delta_{B,i}$ have been optimally fitted to each block.

\Tab{fw100k_medsky_scalar_resid_diffs} lists distributions of both image level differences between coupled blocks, before and after the application of scalar sky offsets.
Uncorrected, the ensemble of coupled blocks have a mean intensity difference of $\sim 1\%$ of the typical sky background intensity.
Scalar sky offsets decrease the differences between overlapping fields to $\sim 0.2\%$.

\begin{figure}[t]
\centering
\includegraphics[width=3.5in]{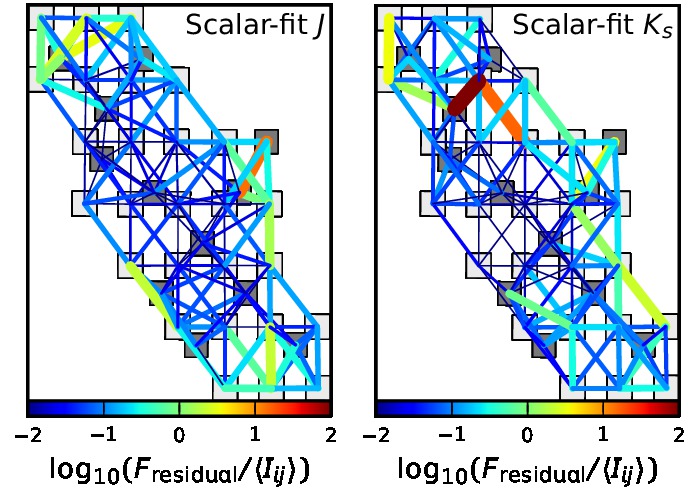}
\caption{Map of residual block-to-block surface brightness differences as a fraction of mean local surface brightness, after scalar fitting.
  This graph mimics the spatial distribution of the 2007B and 2009B WIRCam fields (\Fig{fieldmap}), with the footprints have been exploded to allow room for lines to connect coupled blocks.
}
\label{fig:fw100k_medsky_scalar_resid_sbfrac}
\end{figure}

\begin{figure}[t]
\centering
\includegraphics[width=3.5in]{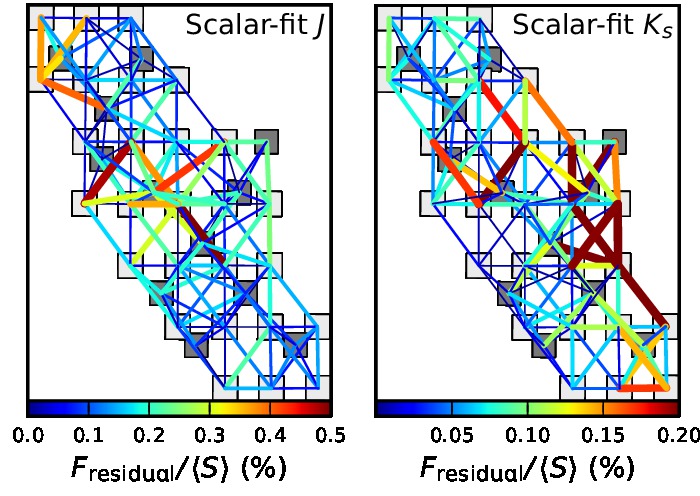}
\caption{Map of residual block-to-block surface brightness differences as a fraction of the mean sky level, after scalar fitting.}
\label{fig:fw100k_medsky_scalar_resid_skyfrac}
\end{figure}

\Fig{fw100k_medsky_scalar_resid_sbfrac} shows the block-to-block residual differences as a fraction of the local surface brightness.
Note that throughout the bright inner disk of M31, block-to-block residuals are negligible compared to the disk signal, at the mosaic periphery ($R\sim 20$~kpc), field-to-field residuals become comparable to, or greater than, the disk surface brightness.
The poor fit is driven primarily by diminishing disk signal, rather than poor convergence of sky offsets.
This can be seen by plotting the magnitude of block-to-block residuals (in units of sky brightness) in \Fig{fw100k_medsky_scalar_resid_skyfrac}.
There, significant residuals are distributed throughout the disk, rather than the low-SB periphery of the mosaic.

\begin{figure}[t]
\centering
\includegraphics[width=3.5in]{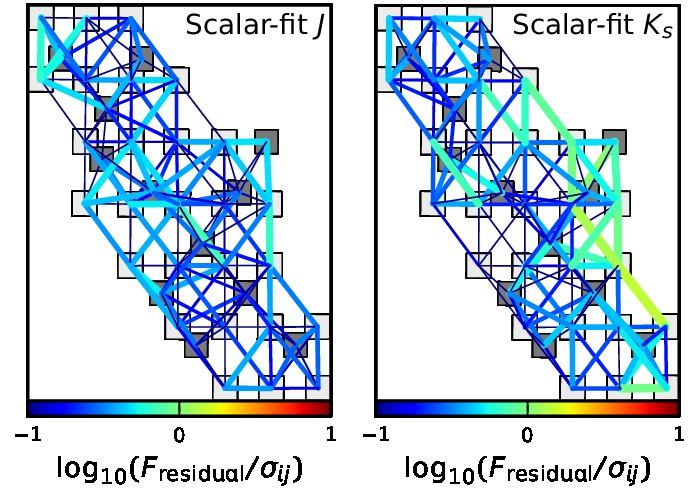}
\caption{Map of residual block-to-block surface brightness differences as a fraction of the standard deviation of the difference image, after scalar fitting.}
\label{fig:fw100k_medsky_scalar_resid_sigmafrac}
\end{figure}

Of course, it is precisely this residual image level difference whose minimization was sought as part of the optimization's objective function (\Eq{objf}).
The inability of scalar sky offset optimization to eliminate residual image differences should not be interpreted as a failure to detect the global minimum; the MSRNM optimization algorithm appears robust in yielding this offset solution set.
Evidence of this can be seen in \Fig{fw100k_medsky_scalar_resid_sigmafrac}, where block-to-block network connections are coloured by the ratio of the residual block-to-block intensity difference to the uncertainty in the block-to-block difference image.
The sky offsets solved by the MSRNM algorithm are within the uncertainties of the difference images themselves; better scalar sky offsets \emph{cannot} be made with the WIRCam blocks that our pipeline has produced.
Nonetheless, block-to-block surface brightness discontinuities are still evident in \Fig{scalar_mosaics}.

An interpretation of this predicament is that sky background residuals are not entirely scalar across WIRCam blocks.
Our discussion in \Sec{skyflatstability} indicates that flat field variability, and sky background variability, can affect the shape of WIRCam fields.

\begin{table}[t]
\centering
\caption[Coupled block differences and residual differences after
scalar sky offsets]{Coupled block intensity differences and residual intensity differences after application of scalar sky offsets: 25th, 50th and 75th percentiles of distribution.
Differences are presented as a percent of the mean sky level seen by observations in each band.
}
\begin{tabular}{lccc}
& \multicolumn{3}{c}{Coupled Block
$\langle I_i - I_j\rangle / \langle I_\mathrm{sky} \rangle$ (\%)} \\
& 25th & 50th & 75th \\
\hline
$J$, uncorrected & 0.47 & 0.91 & 1.70 \\
$J$, scalar offset & 0.05 & 0.09 & 0.18 \\
\hline
$K_s$, uncorrected & 0.42 & 0.90 & 1.43 \\
$K_s$, scalar offset & 0.02 & 0.05 & 0.08 \\
\hline
\end{tabular}
\label{tab:fw100k_medsky_scalar_resid_diffs}
\end{table}

\subsection{The Growth of Sky Offsets in Time and Space}
\label{sec:offsetevo}

How does our sky-target nodding program affect the magnitude of the sky offsets?
This was partially addressed in \Sec{offset_amplitudes}, where both the 2007B and 2009B observing programs resulted in similar net offset distributions (\Tab{offset_hierarchy}).
In this section, we directly map the observed sky offsets as a function of time latency relative to the sky observation.

\begin{figure}[t]
\centering
\includegraphics[width=\columnwidth]{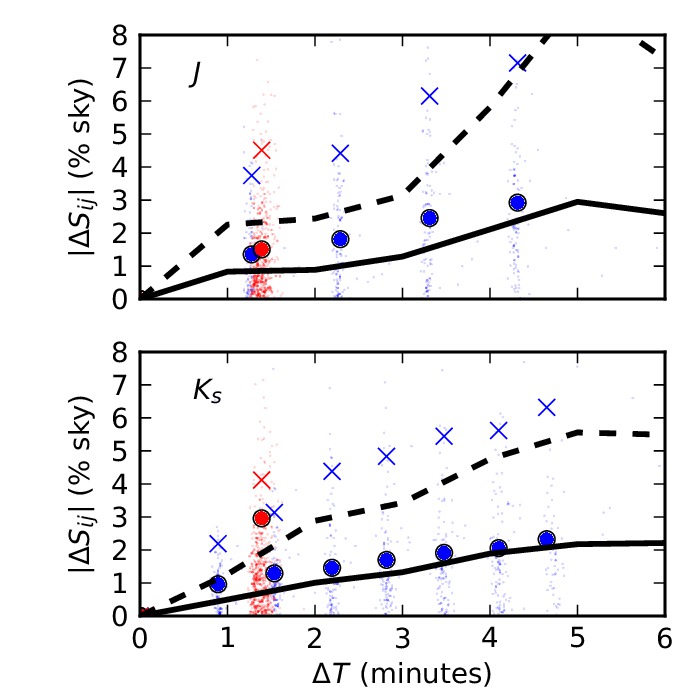}
\caption{Comparison of sky variation distribution functions measured as sky offset amplitudes and between pairs of sky images for $J$ (top) and $K_s$ (bottom) bands.
Mean and 95\% levels of sky variation are plotted as solid and dashed black lines, respectively.
Mean and 95\% levels of sky offsets in the 2007B semester are plotted as large blue circles and X symbols, respectively, while the same for the 2009B semester is plotted as red symbols where sky latency was constrained to 1.2~minutes in both bands.}
\label{fig:nodding_stationary_skyvar_skyfrac}
\end{figure}

First, we must realize that sky offsets are born not only from temporal variability in the sky level (\eg \Fig{skylevel_timeseries}), but also from spatial structure in the NIR skyglow.
Thus as a fiducial, we first build a function of mean sky level variation as a function of time measured at stationary site on the sky (a single sky field), without telescope nodding.
In \Fig{nodding_stationary_skyvar_skyfrac} we plot the mean and 95\% growth of sky level variations as a function of time.
In agreement with \cite{Vaduvescu:2004}, we see a mean sky level variation of $\sim 0.5\%$ in 1~minute in both $J$ and $K_s$ bands.
After 5~minutes, the intrinsic sky level variation typically grows to 2\%.
At worst, we see a sky variation (measured at the 95\% level of the sample distribution) of 5\% in 5 minutes.

Individual points in \Fig{nodding_stationary_skyvar_skyfrac} are net sky offsets of disk images plotted against the time latency to the paired sky sample.
The periodic time structure in \Fig{nodding_stationary_skyvar_skyfrac} is a consequence of the 2007B and 2009B sky-target nodding schemes (see \Tab{obssummary}); indeed the circles and `x' marks in \Fig{nodding_stationary_skyvar_skyfrac} denote the mean and 95\% level of sky variation, respectively, in each cluster.
Recall that all 2009B disk integrations have equal sky sample latency due to the \emph{sky-target-target-sky} nodding pattern.

In both $J$ and $K_s$ bands, we see that nodding the telescope between sky and target \emph{generates} additional sky level uncertainty beyond that expected from strictly temporal sky evolution.
This makes sense in the context of spatial sky variations \citep{Adams:1996}.
As shown in \Fig{nodding_stationary_skyvar_skyfrac}, the process of sky-target nodding can inflate sky variations by 1.5--2 times the sky variability expected at a stationary site on the sky.
On longer time scales, the nodding and stationary sky variance converge, perhaps indicative of the timescales that NIR skyglow structures move across a nodding distance ($1\arcdeg$--$2\arcdeg$ on the sky).

This analysis underscores the challenge of accurately recovering surface brightness in a wide-field NIR mosaic.
Sky-target nodding with CFHT implies typical time latencies of 60--70~seconds, and nodding distances of $1\arcdeg$--$2\arcdeg$.
Both of these elements prevent the true level of the sky on M31's disk, in any single frame, from being known to an accuracy greater than 2\%.

\Fig{nodding_stationary_skyvar_skyfrac} also shows that the 2009B observing strategy of minimizing sky nodding latency would maximize sky level certainty.
The rather shallow slopes of the mean sky variance seen in sky-target nodding demonstrates the modest gain sky certainty by capping sky latency at 1.2 minutes (\ie 2009B) compared to allowing latencies of 5 minutes (\ie 2007B).
This also better explains \Tab{offset_hierarchy}.

\section{Systematic Uncertainties in Surface Brightness Reconstruction}
\label{sec:systematics}

Sky offsets produce a mosaic that is rigorously optimal only in the sense of field-to-field surface brightness continuity---not absolute sky subtraction. In this section, we attempt to gauge the systematic surface brightness error inherent in the sky offset technique.

\subsection{Comparison to \sw{Montage}-fitted images}

\begin{figure}[t]
    \centering
        \includegraphics[width=3.5in]{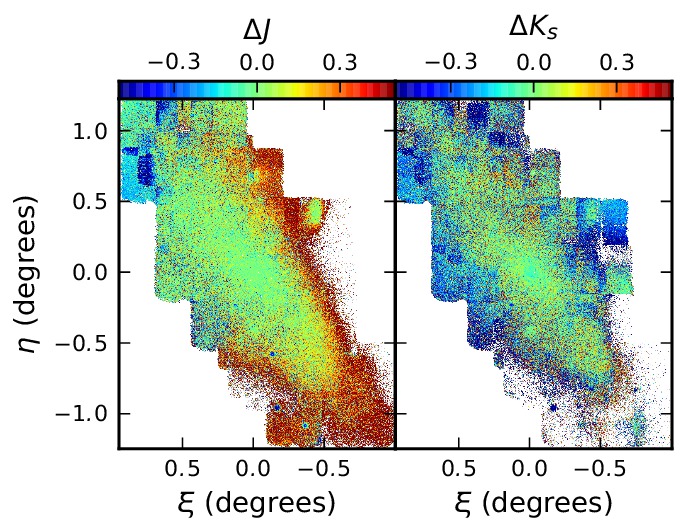}
        \caption{Surface brightness difference maps between our simplex scalar-fit mosaics (\Fig{scalar_mosaics}) and \sw{Montage} scalar-fit mosaics.}
    \label{fig:simplex_scalarmontage_comp}
    \end{figure}

One means of testing for systematic errors in mosaic construction is to compare results from different methods.
Besides the simplex method developed in \Sec{msrnm_algo}--\Sec{hierarchical_algo} and analyzed in detail in Section~\ref{sec:scalaranalysis}, we also tested the \sw{Montage} code that uses an iterative algorithm to solve either scalar or planar sky offsets (see \Sec{montage_algo}).
Figure~\ref{fig:simplex_scalarmontage_comp} shows the surface brightness difference between our simplex solution and the iterative \sw{Montage} mosaic solution assuming scalar sky offsets.
Despite an identical dataset, the two methods yield systematically differences of up to $\sim 0.5$~mag~arcsec$^{-2}$ at 20~kpc, though the solutions are consistent in their treatment of the inner disk.
Although the simplex and \sw{Montage} scalar-offset mosaics \emph{appear} equivalently valid to the eye, a unique and optimal sky offset solution either does not exist, or is extremely difficult for our optimization algorithms to find.

\begin{figure}[t]
\centering
\includegraphics[width=3.5in]{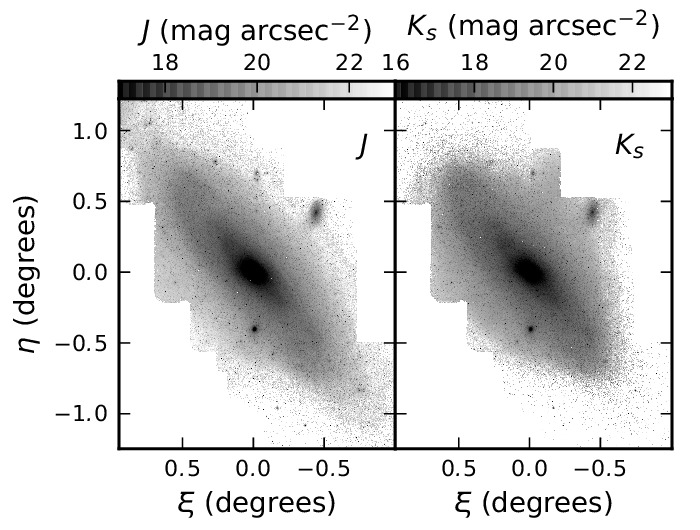}
\caption{Montage-generated $J$ and $K_s$ maps. Compare to the equivalent simplex maps, \Fig{scalar_mosaics}. Pixels with negative surface intensity, after sky offset correction, are masked with white.}
\label{fig:montage_planar_mosaics}
\end{figure}

\sw{Montage} is also capable of fitting planar sky offsets to images, which is a tempting solution to the field-to-field discontinuities that persist between scalar-offset blocks.
The result of planar fitting is shown in \Fig{montage_planar_mosaics}.
We see that planar sky offsets, in this case, do little to improve the mosaics, and indeed, have a dramatic effect on the systematic surface brightness of the mosaic (by more than 1~mag~arcsec$^{-2}$ in the $K_s$ band).
Planar sky offsets may be amenable for observations acquired in long strips (such as 2MASS or SDSS), however, the small and square WIRCam fields do not provide the necessary leverage to prevent systematic error propagation from planar offsets.
\textit{We thus recommend against using planar, or higher-order, sky offsets in wide-field WIRCam mosaics.}

\subsection{Comparison to Spitzer/IRAC Images}

\begin{figure}[t]
\centering
\includegraphics[width=\columnwidth]{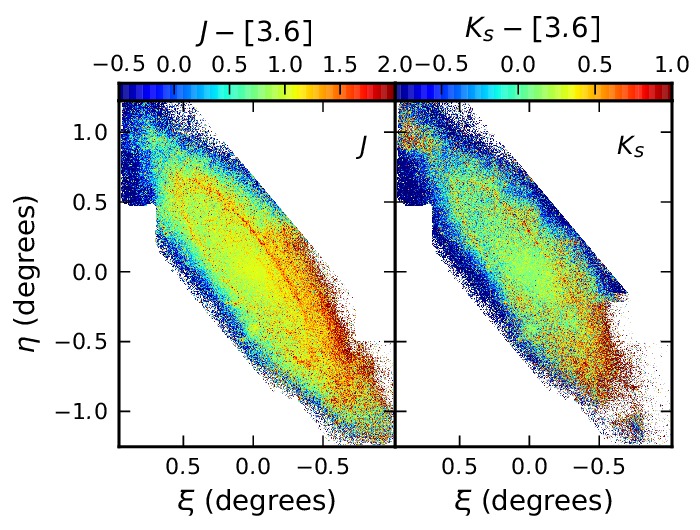}
\caption{Maps of $J-[3.6]$ and $K_s-[3.6]$ surface colour inferred from the simplex scalar-sky fitted WIRCam mosaics and Spitzer/IRAC 3.6~$\mu$m image \citep{Barmby:2006}.
Note that the IRAC map crops the \androids/WIRCam footprint.}
\label{fig:scalar_irac36_sbdiff}
\end{figure}

We also explore systematic uncertainties in our WIRCam mosaics with comparisons against well-calibrated images of M31.
A template for the NIR disk is the 3.6~$\mu$m Spitzer/IRAC map, presented in \cite{Barmby:2006}.
Note that although Spitzer data avoid background subtraction issues caused by the NIR sky, planar sky offsets were used by \citeauthor{Barmby:2006}, though presumably of a smaller magnitude than our WIRCam sky offsets.
In \Fig{scalar_irac36_sbdiff} we compare our simplex scalar-fitted mosaics against the 3.6~$\mu$m image.
Generally the $J-[3.6]$ and $K_s-[3.6]$ colors decrease with disk radius, but increase in the star-forming regions due to hot dust emission.
However both colour maps (coincidentally) become redder in the south western disk beyond the 10~kpc star forming ring.
We interpret this as a systematic over-subtraction of sky in these regions on the order of $\gtrsim 1$ mag~arcsec$^{-2}$.
Evidently, our scalar sky offset mosaics are not systematically reliable beyond the bright disk of M31, toward $R>15$~kpc.

\subsection{Monte Carlo Analysis of Systematic Surface Brightness Uncertainties}
\label{sec:montecarlo}

\begin{figure}[t]
\centering
\includegraphics[width=\columnwidth]{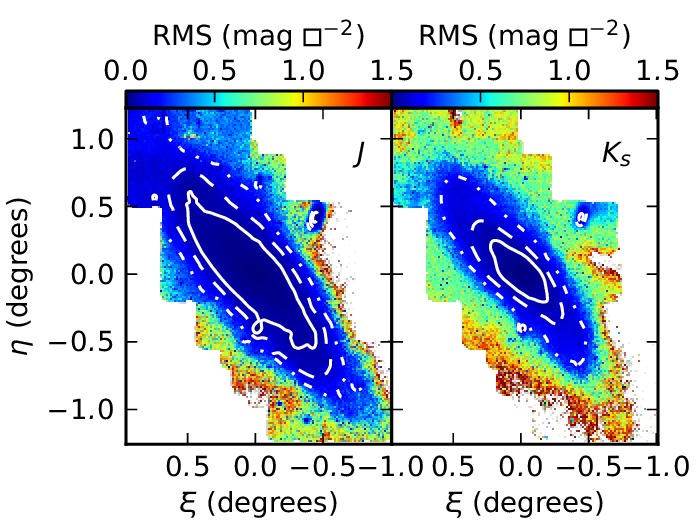}
\caption{Mosaic maps of bootstrap RMS surface brightness in $J$ (left) and $K_s$ (right).
White contours identify RMS levels of 0.05 (solid), 0.1 (dashed) and 0.2 (dash-dot) mag arcsec$^{-2}$.}
\label{fig:bootstrap_sb_rms}
\end{figure}

The difference images presented in the previous section illustrate how the surface brightness reconstructions of identical data can vary depending on the optimization algorithm.
Here we pose a slightly different question: how are reconstructions affected by the initial conditions of sky errors?
That is, given the possible sets of sky level biases affecting the blocks, what is the distribution of surface brightness reconstructions?
We answer this with a realistic Monte Carlo analysis.

A Monte Carlo (MC) realization is generated by perturbing the surface brightness of the corrected blocks with a sky error drawn (with replacement) from the ensemble of block sky offsets observed in the original mosaic (\Fig{scalar_mosaics}).
Using the scalar-sky fitting procedure, sky offsets  are optimized against the known sky perturbations; 100 such realizations are made to compile an ensemble of mosaics in both bands.

Figure~\ref{fig:bootstrap_sb_rms} shows the RMS deviation of MC mosaic surface brightness against the original scalar-fitted mosaics.
Reconstructed surface brightness in the outer disk can vary by $\sim 1$~mag~arcsec$^{-2}$, consistent with colour biases in the $J-[3.6]$ and $K_s-[3.6]$ maps.

We can ultimately understand the source of surface brightness by examining the standard deviations in the residual between expected and realized sky offsets in each Monte Carlo iteration.
This residual dispersion is 0.15\% of the $J$-sky (0.17\% of the $K_s$ sky); we find this dispersion to be constant across all fields in the mosaics.
If mosaic surface brightness uncertainty is caused by flexure in the mosaic---where blocks on the mosaic periphery are forced to conform to the surface brightness of more central and tightly coupled blocks---then outer blocks would have higher offset dispersion.
This is not the case.

Rather than mosaic flexure, a better model for \Fig{bootstrap_sb_rms} involves uncertainties in the \textit{post priori} adjustment for zero net offset (\Eq{netzero}).
Since block sky offsets have approximately Gaussian distributions with dispersions given in \Tab{offset_hierarchy}, the uncertainty in the net offset correction is simply $\sigma(\mathrm{block})/\sqrt{n_\mathrm{blocks}}$, where $n_\mathrm{blocks}=39$ in the combined 2007B and 2009B mosaic.
Given that $\sigma_{\Delta_B}\sim 1\%$, the expected uncertainty in the net offset correction is 0.16\%: in perfect correspondence to the observed mosaic uncertainty.
The dominant source of uncertainty shown in the MC simulations, \Fig{bootstrap_sb_rms}, is the use of an arithmetic mean of offsets to set an absolute zeropoint, not flexure or uncertainty in the network of offsets.
This suggests that external zeropoints could be very useful in replacing \Eq{netzero}.
Since no absolutely-calibrated NIR photometry of M31's surface brightness exists, we will discuss a method using panchromatic resolved stellar populations in a future work.

\section{Accuracy of Surface Brightness Shapes Across WIRCam Frames}
\label{sec:skyflatstability}

Our analysis of the WIRCam M31 mosaics has thus far focused on macroscopic surface brightness accuracy.
Here we examine the accuracy of surface brightness shapes in individual WIRCam frames (and ultimately, blocks).
Such shape biases give rise to discontinuities between blocks in our optimized mosaics (\Fig{scalar_mosaics}), and reflect high-order irregularities in block surface brightness shape that cannot be corrected with either scalar (zeroth-order) or even planar (first-order) sky offsets.

Block shape accuracy is affected by two stages of our reduction pipeline: first, in the \emph{proportional} corrections of flat-fielding, and second, in the \emph{additive} corrections of median sky subtraction.
The accuracy and effectiveness of both calibrations is limited by spatial and temporal variations in the NIR sky itself (see discussion in \Sec{intro}).
In this section, we attempt to deconvolve the scales of proportional and additive surface brightness biases, and ultimately establish an empirical upper limit on the edge-to-edge surface brightness accuracy seen in our WIRCam program.

\subsection{Evolution of Real Time Sky Flats} \label{sec:flatevo}

In producing a NIR sky flat, we assert that the mean shape of the NIR sky over a timespan is flat.
By maximizing the time window we can marginalize over as many sky shapes as possible to produce an unbiased skyflat.
This notion is embodied in the \texttt{QRUN} sky flats that combine hundreds of sky shapes captured across several days.
However, this also assumes that WIRCam is a \emph{stable} detector over the period of several days; our real-time \texttt{FW100K} sky flats, however, assume that WIRCam is unstable over periods of 30-minutes.

\begin{figure*}[t]
\centering
\includegraphics[width=\textwidth]{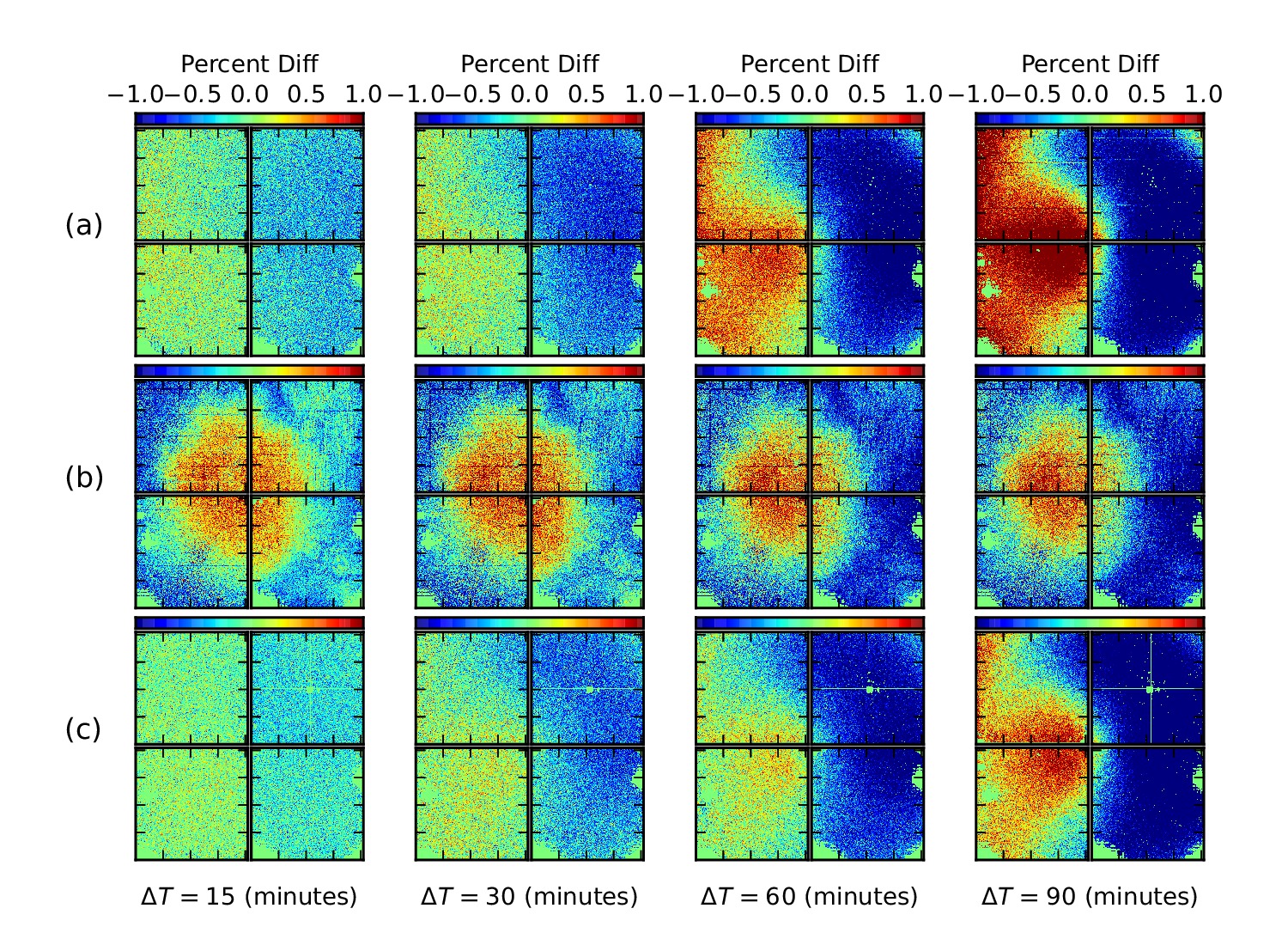}
\caption{Spatio-temporal evolution of \texttt{FW100K} sky flats, shown as percent difference of the flat fields 15, 30, 60 and 90 minutes after the initial sky flat of the night.
Sky flat evolution on three nights is shown: (a) 2009B $K_s$ observing block with the telescope staring at a sky field, (b) 2009B $K_s$ observing block with regular sky-target nodding, and (c) a 2007B observing block with sky-target nodding.}
\label{fig:fw100k_movie}
\end{figure*}

A simple test of WIRCam's flatfield stability is to monitor how the real-time \texttt{FW100K} skyflats evolve on scales of hours.
In \Fig{fw100k_movie} we show percent difference images between \texttt{FW100K} sky flats made at intervals of 15, 30, 60 and 90 minutes after an initial sky flat.
After just 30~minutes, the shape of the \texttt{FW100K} skyflats deviates by 0.5\% from the initial flat field shape.
By 60 minutes, the deviation exceeds 1\%.
Evidently, the WIRCam flat field function is stable on timescales less than 30 minutes---much less than a queue run.

Nonetheless, the spatio-temporal evolution takes many forms.
In some cases (\Fig{fw100k_movie}b) the flat field deviations are axisymmetric, while in others there is a distinct East-West pattern (\Fig{fw100k_movie}ac).
We interpret these as instabilities in the WIRCam illumination function on the scale of minutes.

\begin{figure}[t]
\centering
\includegraphics[width=\columnwidth]{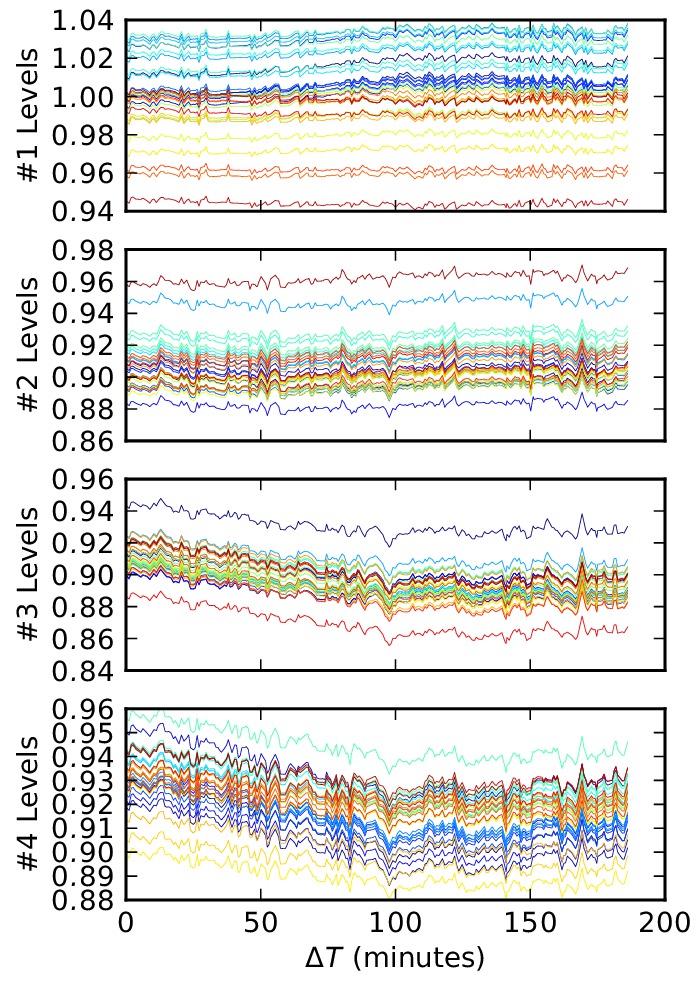}
\caption{Time evolution of the mean flat field level of amplifier bands in the four WIRCam detectors in \texttt{FW100K} sky flats made over three hours.
Amplifier bands are colour-coded according to their order on the array: red lines at the bottom, green in the middle, and blue at the top.
Although the levels of individual amplifiers differ by 10\%, their order is consistent, indicating that amplifier gain is extremely stable.
The jitter is due to detector-to-detector zeropoint normalization (\Sec{flatbuilding}) uncertainty, or measurement biases.}
\label{fig:fw100k_globalamp_timeseries_55136_Ks}
\end{figure}

An alternative interpretation is that these sky flat deviations are instabilities in the WIRCam detector electronics.
The dominant macroscopic electronic feature in WIRCam flat fields are the amplifier bands.
Each WIRCam detector is divided into 32 horizontal bands (each 64 pixels high) that are read out into independent amplifiers.
These amplifiers have gains that result in levels that differ by 10\% in flat field images.
Still, these different gains appear stable: over the course of an observing block, the mean flat field level of each amplifier band evolved by less than 0.1\% relative to other amplifiers (see \Fig{fw100k_globalamp_timeseries_55136_Ks}) over three hours.
Indeed, the amplifier band signature is absent from the \texttt{FW100K} skyflat difference images (\Fig{fw100k_movie}).\footnote{Quite unlike \Fig{domeflatratio} that compared dome flat and \texttt{QRUN} sky flat shapes.}
Thus sky flat evolution does appear driven by changes in the large scale detector illumination function, not electronic instabilities.

\subsection{Shapes of Median Sky Frames}
\label{sec:medianskyshapes}

\begin{figure}[t]
\centering
\includegraphics[width=\columnwidth]{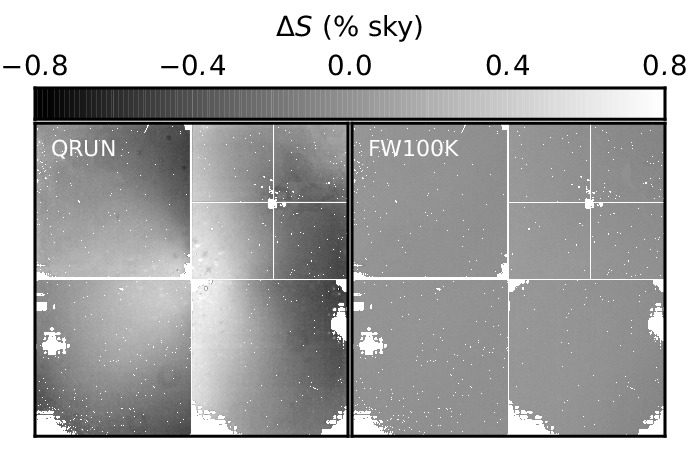}
\caption{Comparison of a typical median sky frames in $2\times 2$ WIRCam array in the $K_s$-band from images calibrated with \texttt{QRUN} (left) and \texttt{FW100K} (right) sky flats.
Amplitudes of shapes in these median sky images are plotted as a fraction of the mean sky brightness.
While the real-time sky flat produces very flat median sky images, \texttt{QRUN}-calibrated data produce sky biases on the order of 1\% of the NIR sky level.
}
\label{fig:median_sky_images}
\end{figure}

Another test of sky flats is their ability to produce an unbiased sky background, up to the level of intrinsic sky variations.
This test can be made by examining the median sky frames (\Sec{mediansky}) produced by \texttt{QRUN} and \texttt{FW100K} sky flats, as shown in \Fig{median_sky_images}.
\texttt{QRUN} sky flats clearly produce biased sky shapes, evidently caused by changes in the CFHT/WIRCam optical path and electronics over a queue run, rather than intrinsic variations in the sky itself.

\begin{figure}[t]
\centering
\includegraphics[width=\columnwidth]{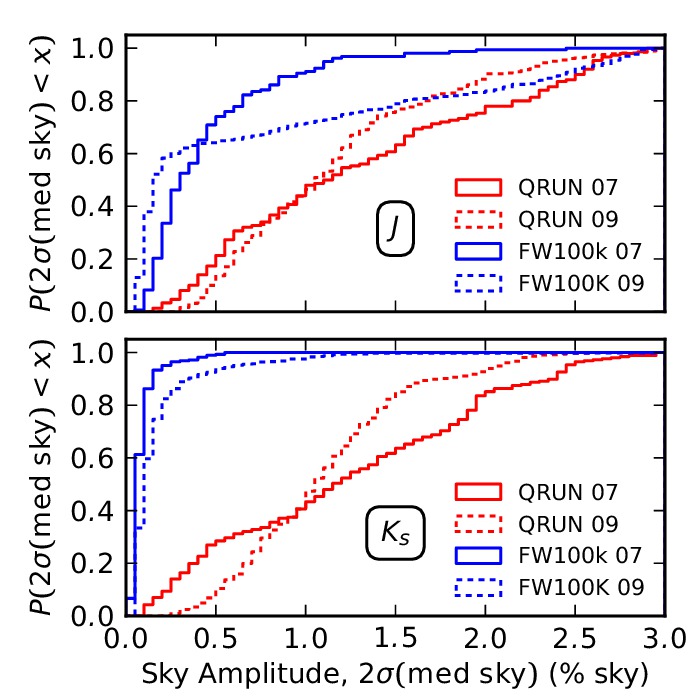}
\caption{Cumulative distribution function of sky level amplitudes across median sky images processed with \texttt{QRUN} (black) and \texttt{FW100K} (blue) sky flats for the $J$ (top) and $K_s$ (bottom) bands. Real-time (\texttt{FW100k}) sky flats produce much flatter median sky frames, with expected shape amplitudes of 0.3\% ($J$) to 0.1\% ($K_s$) of the NIR sky level, while the expected amplitude of \texttt{QRUN}-flat processed sky images is 0.5\% of the sky level, and as high as 3\% of the sky level.
}
\label{fig:mediansky_amplitude}
\end{figure}

To quantitatively compare the bias of \texttt{QRUN} versus \texttt{FW100K} sky flats, we measure the amplitude of shapes across the $10\arcmin \times 10\arcmin$ WIRCam frame as the 2-standard deviation interval (95\%) of each median sky image's pixel distribution: $2 \sigma(\mathrm{med~sky})$.
These distributions of sky shape amplitude are plotted in \Fig{mediansky_amplitude}.
As expected from \Fig{median_sky_images}, \texttt{QRUN} median sky frames have typical biases of 1\% of the sky level, while sky frames from real-time \texttt{FW100K} sky flats are much flatter, $\lesssim 0.3\%$ of the sky level, and more consistently so.

Although \texttt{FW100k} sky flats are better than \texttt{QRUN} flats, we can still wonder if the 0.3\% flatness limit of median sky images is a consequence of intrinsic sky shapes or due to limits in the WIRCam flat field accuracy.
Recall that median sky images are composed of five sky frames taken closest to a disk frame.
In 2007B, all sky frames were sampled from the same coordinate on the sky, and span a 12~minute window covering sky integrations taken before and after a disk image (for both $J$ and $K_s$ sky-target nods).
In 2009B, sky frames were sampled from randomly chosen sites along the sky field ring (\Fig{fieldmap}) with a window typically spanning 15 minutes.
Thus both 2007B and 2009B median sky images span similar time windows, although the 2009B strategy attempts to marginalize over five distinct sites  on the sky (and thus sky shapes) while 2007B median sky images do not.
If the spatial skyglow pattern and WIRCam flatness function were stationary, we expect the instantaneous shape of the sky (in the sky field) to be captured in the 2007B median sky frames, while the 2009B sky frames should marginalize over five random sky shapes and thus be flatter by up to a factor of $1/\sqrt{5}$ (for a Gaussian shape distribution).
That this \emph{does not} occur indicates either that (a) sky shapes are correlated over 15~minutes and 3\arcdeg\ of sky, (b) skyglow patterns have sufficient temporal variability to be effectively uncorrelated over 15~minute windows across a WIRCam frame so that both observing patterns marginalize over sky shapes equally, or (c) the flatness of median sky images is limited at the 0.3\% level by background variations associated with WIRCam itself over 15~minutes.
Wide-field movies of the NIR sky \citep{Adams:1996} suggest option (a) to be false.
From this test alone, then, we cannot distinguish between instrumental background variability or stochasticity in the sky as the cause of the 0.3\% sky shape amplitudes seen by WIRCam median sky images.
Further, this test cannot distinguish between proportional instrumental variability in the flat field (\eg \Fig{fw100k_movie}) or additive background variability \citep[as seen in the CFHT-IR camera,][]{Vaduvescu:2004}.

\subsection{Frame residuals shapes}
\label{sec:frameblockresiduals}

In previous sections, we have shown that real-time \texttt{FW100K} sky flats are preferred over broader-baseline \texttt{QRUN} sky flats since they produce systematically flatter median sky images, and do seem to track real evolution in the WIRCam flat field function on scales of 30 minutes.
Yet it is difficult to measure the absolute accuracy of real-time sky flats since flat field bias cannot be distinguished from additive background stochasticity (from sky or instrument) when simply observing the shape of the sky background.
If we examine signal- (not sky-) dominated fields, flat fielding biases should grow in comparison to sky shape biases.

\begin{figure*}[p]
\centering
\includegraphics[width=\textwidth]{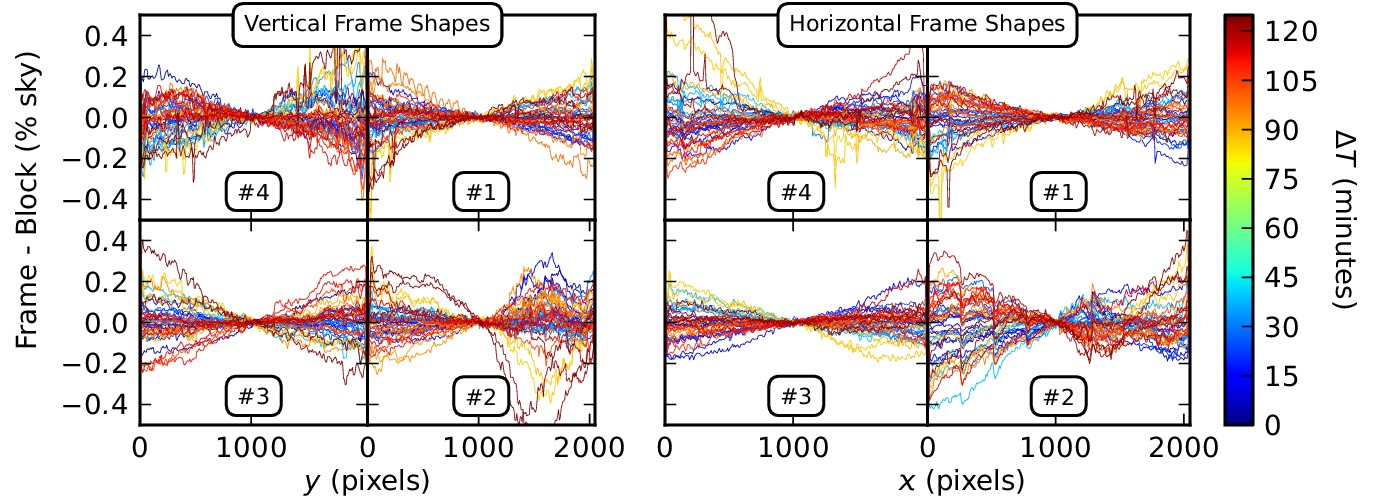}
\caption{Residual shapes of individual \texttt{FW100k} sky flat-processed WIRCam integrations of the M31 disk to the median (mosaic) shape for the field M31-37, $K_s$-band, observed in the 2009B semester.
Residuals have been marginalized across the $x$ (left) and ($y$) (right) axes to provide 1D views.
Axes match WIRCam's $2\times2$ detector footprint.
Individual integrations are coloured by their time after the first disk integration.
The centre of M31 is located in the lower-right detector (\#2); surface brightness bias in these regions betray the presence of flat field bias. Lower surface brightness regions are dominated by shape variations on the order of $\pm 0.2\%$ of sky, interpreted as additive uncertainties associated either with the detector, skyglow, or both.}
\label{fig:frame_residuals_M31-37_Ks_fw100k_medsky}
\end{figure*}

\begin{figure*}[p]
\centering
\includegraphics[width=\textwidth]{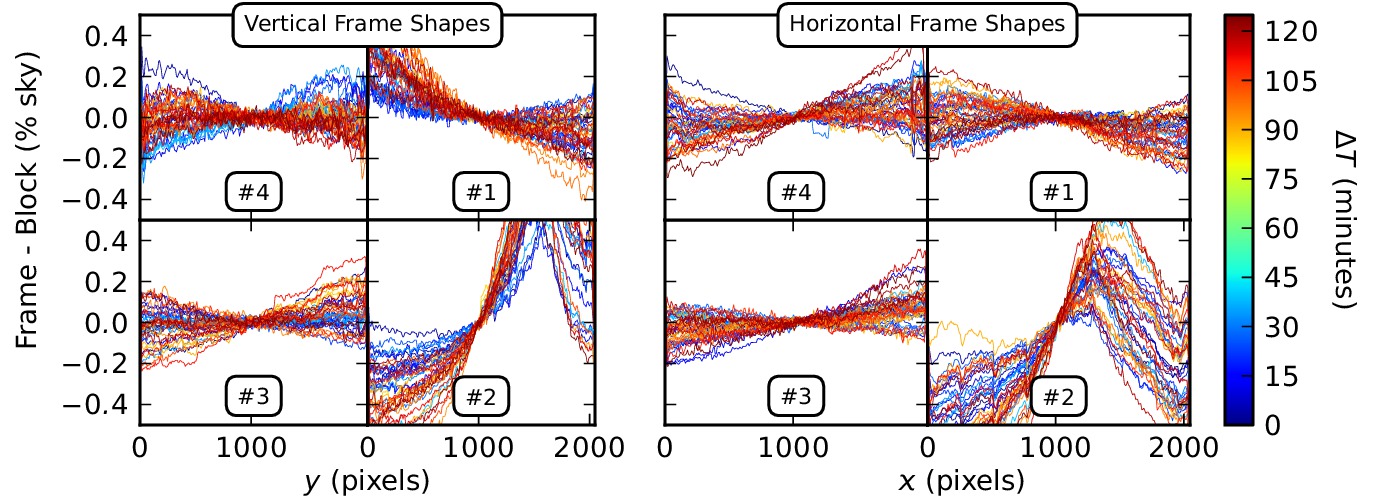}
\caption{Residual shapes of individual of the M31-37 $K_s$ field, processed with \texttt{QRUN} sky flats.
Compared to \Fig{frame_residuals_M31-37_Ks_fw100k_medsky}, \texttt{QRUN} sky flats clearly do not capture flat field evolution that occurs over the course of 90-minutes, yielding systematically evolving realizations of bulge-dominated surface brightness in detector \#2 (lower-right). In the more sky-dominated regions of the image (detector \#4), \texttt{QRUN} sky flats produce images with similar stability to \texttt{FW100K} sky flats, indicating that the limit of additive uncertainties associated with sky or instrumental background variations is reached here.}
\label{fig:frame_residuals_M31-37_Ks_QRUN}
\end{figure*}

Our 2009B observations of field M31-37 in the $K_s$ band are ideal for this experiment: a single detector in that field covers the core of M31, and observations were taken into two blocks, covering a total window of 2~hours (most blocks for this program are observed by the CFHT queue in a half hour).
Both the high surface brightness and wide time baseline in this field should highlight flat field bias and variation.
In Figures~\ref{fig:frame_residuals_M31-37_Ks_fw100k_medsky} and~\ref{fig:frame_residuals_M31-37_Ks_QRUN} we show the residual shapes of individual WIRCam frames against the median shape of the mosaic, given \texttt{FW100K} and \texttt{QRUN} sky flattening, respectively, in field M31-37 in the $K_s$ band.
That is, we produce difference images between each WIRCam frame and the block mosaic.
To analyze the shapes of these difference images we then marginalize the difference images along rows (left side of \Fig{frame_residuals_M31-37_Ks_fw100k_medsky}), and columns (right side of \Fig{frame_residuals_M31-37_Ks_fw100k_medsky}).
Note that these marginalization are done for each detector in the $2\times2$ WIRCam array; the core of M31 resides in detector \#2 (lower-right).
In that high surface brightness region, there are strong surface brightness residuals that clearly point out flaws in the flat field itself.
\texttt{FW100K} sky flats (\Fig{frame_residuals_M31-37_Ks_fw100k_medsky}) WIRCam frames at the core of M31 can vary in surface brightness by $\pm 0.5\%$; with \texttt{QRUN} sky flats these residuals are much larger, nearly $\pm 1\%$ of the $K_s$-band sky brightness.
The time scale residual evolution clearly matches that of flat field evolution indicated in \Fig{fw100k_movie}.
Despite limited accuracy, the \texttt{FW100K} sky flats do appear to track the real-time evolution of the WIRCam detector and produce more consistent frame shapes.
Although all frames in \Fig{frame_residuals_M31-37_Ks_QRUN} are processed with the same \texttt{QRUN} sky flat, the intrinsic flat field function of WIRCam has evolved over the span of two hours to create frame-to-frame shape variations nearly $\pm 1\%$ of the $K_s$-band sky brightness at the center of M31.

It is useful to contrast the frame shape residuals seen in detector \#2 with those in other detectors, where the disk surface brightness is lower.
There, both \texttt{FW100K} (\Fig{frame_residuals_M31-37_Ks_fw100k_medsky}) and \texttt{QRUN} (\Fig{frame_residuals_M31-37_Ks_QRUN}) show similar residual distributions, on the order of $\lesssim 0.2$\% of the NIR sky brightness.
Further, the results are not monotonically varying in time, as they are in detector \#2, and indeed appear to vary essentially randomly.
We interpret this frame residual behaviour as being caused by random additive background processes, distinct from flat field biases that are proportional to surface brightness.

\subsubsection{Distributions of frame shape residuals}
\label{sec:frameblockresidualhist}

\begin{figure}[t]
\centering
\includegraphics[width=\columnwidth]{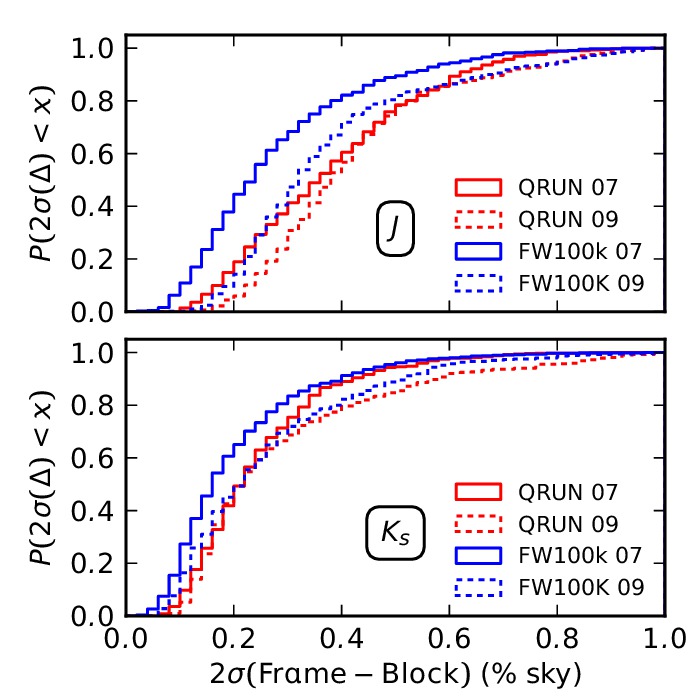}
\caption{Cumulative distributions of scalar difference amplitudes between individual frames and blocks in the $J$ (top) and $K_s$ (bottom) mosaics, measured as a dispersion of pixel differences at the 95\% level.
Whether processed with \texttt{QRUN} (red) or \texttt{FW100k} (blue) sky flats, or observed in 2007B (solid lines) or 2009B (dotted lines), the residual amplitude differences between frames and blocks are similarly distributed.
The expected amplitude difference is $0.3\%$ of the $J$ sky brightness ($0.2\%$ in $K_s$).
}
\label{fig:frame_diffs_skyfrac}
\end{figure}

We now extend the previous analysis analysis across the entire dataset.
In \Fig{frame_diffs_skyfrac} we plot the distribution of frame-block residual shape amplitudes, measured at the 95\% difference interval.
In essence, this measures the consistency of imaging the shape of each M31 block.
As in our test of median sky frame flatness (\Sec{medianskyshapes}), we see that the consistency of frame shapes is $\sim 0.3\%$ of the sky level.
This result is seen for both \texttt{QRUN} and \texttt{FW100K} sky flat pipelines and for 2007B and 2009B observing schemes, agreeing with our observation in \Sec{frameblockresiduals} that in background dominated regimes (as most of our blocks are) frame shape consistency is \emph{not} correlated with flat field bias.
Rather, we interpret \Fig{frame_diffs_skyfrac} as measuring the amplitudes of \emph{additive} stochastic background shapes originating either from the sky, or associated with the instrumentation itself.
Effectively, \Fig{frame_diffs_skyfrac} illustrates the \emph{flatness limit} of WIRCam frames observed with large sky-target nods, sky flat fielding, and median sky subtraction.

\subsection{Section Summary}
\label{sec:shapeconclusions}

We summarize our findings on the accuracy of surface brightness shapes reproduced by WIRCam in a sky-target nodding observing program:

\begin{enumerate}
  \item Surface brightness perturbations can be decomposed into multiplicative processes (flat field biases) and additive processes (stochastic sky and instrumental backgrounds) by observing residual shapes of individual WIRCam frames again the median M31 mosaic in high and low surface brightness regimes; we find evidence for both processes occurring.
  \item The WIRCam flat field function can vary by approximately 1\% across a $10\arcmin \times 10\arcmin$ frame in 1~hour. Building sky flats concurrently with observations is necessary to minimize systematic surface brightness bias. This, however, is only a concern in signal-dominated pixels (such as galaxy centres, or point sources); otherwise median sky subtraction is effective at hiding flat field bias in low-surface brightness features.
  \item In low surface brightness regimes, we observe stochastic variations of $\pm 0.2$\% in the sky amplitude. These background shapes cannot be removed with median sky frame subtraction (which itself has shape amplitudes on the order of $0.3\%$ of the NIR sky level). We find this to be the limiting surface brightness accuracy across a $10\arcmin \times 10\arcmin$ WIRCam frame.
\end{enumerate}

\section{Conclusions}
\label{sec:conclusions}

We have presented near-infrared ($J$ and $K_s$) images of M31's entire bulge and disk with CFHT/WIRCam.
These maps surpass the 2MASS \citep{Beaton:2007} and Spitzer \citep{Barmby:2006} mosaics with superior resolution that permits the identification of individual stars throughout M31's mid and outer disk.
The dataset is also complementary to the HST/WF3 PHAT survey \citep{Dalcanton:2012} by providing complete coverage of M31's disk within $R=22$~kpc, and by offering a broader NIR colour baseline ($J-K_s$) than is offered by WF3 (approximately $J-H$).
NIR mosaics of M31 have crucial applications for studies of the nearly attenuation-free stellar structure of our nearest spiral neighbour, and for tests of stellar population synthesis models in NIR regimes.

Our focus in this paper has been the establishment of procedures for accurately recovering the NIR surface brightness across 3~sq.~deg.\ of the M31 disk using a sky-target nodding observing strategy with WIRCam on CFHT\@.
We have compared two different observing methods to study the effects of sky target nodding cadences and patterns on sky subtraction uncertainties.
We have also developed and tested our WIRCam pipeline for flat fielding, zeropoint estimation, median sky subtraction, and sky offset optimization.

We find that our NIR SB reconstruction is limited in two regimes: large scale reconstruction of surface brightness with sky offsets, and surface brightness \emph{shape} uncertainties across individual WIRCam images.
On a large scale, the necessity of nodding between sky and target limits our direct knowledge of the sky level on the disk by $\gtrsim 2$\% of the sky level.
Strictly minimizing latency between sky and disk integrations (as in the 2009B program) provides only limited improvement in our knowledge of the sky level on the disk because of the overhead in nodding the telescope, and spatial structure in the NIR skyglow itself.
Sky offset optimization is successful in reducing block-to-block surface brightness differences to $<0.1$\% of the sky level.
Given a realized set of blocks, our optimization algorithm reliably finds a consistent solution, so any errors in surface brightness shape across the mosaic are caused by errors in the shapes of individual blocks (see below).
There is, however, an uncertainty in our net zero offset, of order $\sim \sigma_{\Delta_B} / \sqrt{N_\mathrm{blocks}}$; 0.16\% of the sky level.
This zeropoint can ultimately be established using resolved stellar populations, the subject of a forthcoming \androids\ study.

The shape of a WIRCam frame can be affected by both flat field uncertainties and additive background uncertainties.
First, we have discovered that the WIRCam flat field function can change by 1\% in 60~minutes.
This effect is predominantly influential in the signal-dominated regime of the M31 bulge; though it is also significant for resolved stellar photometry.
We thus find that constructing real-time sky flats is essential for calibrating WIRCam images.
In sky-dominated regimes, the 2D SB shapes of individual WIRCam frames ($20\arcmin\times20\arcmin$) are uncertain by 0.2\% of the sky intensity, centre-to-edge.
These background fluctuations present a lower limit in the surface brightness uncertainty across a WIRCam frame since they appear to be caused by skyglow variations in target images that cannot be fully corrected with median sky subtraction.

We now summarize our analysis of the data taking and reduction methods developed in this work, and in doing so, formulate a set of best practices for similar wide-field NIR surveys employing sky-target nodding.

\subsection{Recommendations for Conducting a Wide-field NIR Survey with Sky-Target Nodding}

We first recommend that the sky-target nodding cadence be set to effectively build real-time sky flats, rather than simply track sky level evolution.
Such a program would involve sufficient sky frames to build a sky flat within a window of 20--30 minutes, where the sky is observed in several epochs at different locations in a sampling ring to minimize biases.

In the $K_s$ band this objective is efficiently achieved, since the mean sky flux on a WIRCam pixel is $450 \pm 80$ ADU~$\mathrm{s^{-1}}$.
Given $T_\mathrm{exp}=25$~s, a $\mathrm{[S^3T^6]^3}$ program yields the necessary 9 sky integrations within 20~minutes and a maximum sky-target latency of 2~minutes.
Each sky flat would be built from observations at three sites along the sky ring.

Lower sky flux in the $J$ band ($120 \pm 30$~ADU~s$^{-1}$) requires additional sky integration  to achieve comparable sky flat $S/N$ as the $K_s$ band.
Given $T_\mathrm{exp}=45$~s, a $\mathrm{[S^4T^4]^5}$ program yields the necessary 20 sky integrations in $\sim 40$ minutes, with a maximum sky-target latency of 2.2~minutes.

Sky offsets optimization is aided by having more independent blocks covering the target, since our net zero offset assertion is uncertain by $\sigma_{\Delta_B} / \sqrt{N_\mathrm{blocks}}$.
Given that $\sigma_B$ cannot be reduced, increasing the number of \emph{independent} blocks (observed hours or even a night apart to decouple sky and instrumental biases) is the most reliable way to establish the absolute surface brightness accuracy of the mosaic.
Since sky offsets are further biased by any shape errors in blocks (realized as our inability to diminish block-to-block offsets below $\sim 0.1\%$ of sky brightness), we propose that blocks be interlaced by 50\% (so that one detector completely overlaps a detector from an adjacent blocks).
This interlacing pattern would thus enable the marginalization of shape errors across the entire detector frame.
By doubling the number of blocks, each with individually halved exposure times, the mosaic could be reproduced with an equivalent net integration time.

\bigskip We thank Loic Albert and Karun Thanjavur, previously from CFHT, for their help with WIRCam I'iwi data products and procedures and the CFHT staff for many informative discussions and their diligence in performing the queue service observations. We are grateful for the Spitzer 3.6~$\mu$m mapping provided by Pauline Barmby (University of Western Ontario).
J.S. and S.C. acknowledge support through respective Graduate Scholarship and Discovery grants from the Natural Sciences and Engineering Research Council of Canada.
M.M is supported by NASA through a Hubble Fellowship grant HST-HF51308.01-A awarded by the Space Telescope Science Institute, which is operated by the Association of Universities for Research in Astronomy, Inc., for NASA, under contract NAS 5-26555.
This publication makes use of data products from the Two Micron All Sky Survey, which is a joint project of the University of Massachusetts and the Infrared Processing and Analysis Center/California Institute of Technology, funded by the National Aeronautics and Space Administration and the National Science Foundation.
This work is based on observations obtained with WIRCam, a joint project of CFHT, Taiwan, Korea, Canada, France, at the Canada-France-Hawaii Telescope (CFHT) which is operated by the National Research Council (NRC) of Canada, the Institute National des Sciences de l'Univers of the Centre National de la Recherche Scientifique of France, and the University of Hawaii. 
{\it Facilities:} \facility{CFHT (WIRCam)}, \facility{FLWO:2MASS}.

\bibliography{master}

\end{document}